\documentclass[%
 reprint,
superscriptaddress,
showkeys,
nofootinbib,
amsmath,amssymb,
aps,
pre, 
]{revtex4-2}

\usepackage{lineno}
\usepackage{graphicx}
\usepackage[caption=false]{subfig} %?
\usepackage{chngcntr} % To reset figure numbering in the appendix
\usepackage[normalem]{ulem}
\usepackage{amsmath}
\usepackage[hypertexnames=false]{hyperref} % option should solve issue with hyperlinks not reaching SI figures
\hypersetup{
    colorlinks=true,
    linkcolor=blue,
    filecolor=magenta,      
    urlcolor=blue,
    citecolor = purple,
    pdfpagemode=FullScreen,
    }
\usepackage{array}
\usepackage{booktabs}
\usepackage{float}
\usepackage{amsmath}
\usepackage{amsfonts}
\usepackage{amssymb}
\usepackage{csquotes}
\usepackage{float}
\usepackage{ulem}
\newcommand{\ee}{\end{equation}} 
\newcommand{\be}{\begin{equation}}

\usepackage[mathscr]{euscript}  
\usepackage{xcolor}  

\usepackage{tcolorbox}
\usepackage{comment}
\usepackage{float}
\usepackage{graphicx}

\makeatletter
\newsavebox{\@brx}
\newcommand{\llangle}[1][]{\savebox{\@brx}{\(\m@th{#1\langle}\)}%
  \mathopen{\copy\@brx\kern-0.5\wd\@brx\usebox{\@brx}}}
\newcommand{\rrangle}[1][]{\savebox{\@brx}{\(\m@th{#1\rangle}\)}%
  \mathclose{\copy\@brx\kern-0.5\wd\@brx\usebox{\@brx}}}
\makeatother

\newcommand {\resub}[1]{{\color {black} #1}} 
\newcommand {\finalresub}[1]{{\color {black} #1}} 

\begin{document} 

\preprint{APS/123-QED}

\title{Emergence of ecological structure and species rarity from fluctuating metabolic strategies}
\author{Davide Zanchetta}
\email{davide.zanchetta@phd.unipd.it}
\affiliation{Dipartimento di Fisica e Astronomia “Galileo Galilei”, Universit\`a degli studi di Padova, via Marzolo 8, 35131 Padova, Italy}
\author{Deepak Gupta}
\email{phydeepak.gupta@gmail.com}
\affiliation{Institut für Theoretische Physik, Technische Universität Berlin, Hardenbergstraße 36, D-10623 Berlin, Germany}
\author{Sofia Moschin}
\email{sofia.moschin@studenti.unipd.it}
\affiliation{Dipartimento di Fisica e Astronomia “Galileo Galilei”, Universit\`a degli studi di Padova, via Marzolo 8, 35131 Padova, Italy}
\author{Samir Suweis}
\email{samir.suweis@unipd.it}
\affiliation{Dipartimento di Fisica e Astronomia “Galileo Galilei”, Universit\`a degli studi di Padova, via Marzolo 8, 35131 Padova, Italy}
\affiliation{INFN, Sezione di Padova, via Marzolo 8, Padova, Italy - 35131}%
\affiliation{Padova Neuroscience Center, University of Padova, Padova, Italy}
\author{Amos Maritan}
\email{amos.maritan@unipd.it}
\affiliation{Dipartimento di Fisica e Astronomia “Galileo Galilei”, Universit\`a degli studi di Padova, via Marzolo 8, 35131 Padova, Italy}
\affiliation{INFN, Sezione di Padova, via Marzolo 8, Padova, Italy - 35131}%
\affiliation{National Biodiversity Future Center, Piazza Marina 61, 90133 Palermo, Italy}
\author{Sandro Azaele}
\email{sandro.azaele@unipd.it}
\affiliation{Dipartimento di Fisica e Astronomia “Galileo Galilei”, Universit\`a degli studi di Padova, via Marzolo 8, 35131 Padova, Italy}
\affiliation{INFN, Sezione di Padova, via Marzolo 8, Padova, Italy - 35131}%
\affiliation{National Biodiversity Future Center, Piazza Marina 61, 90133 Palermo, Italy}

\begin{abstract}
\resub{Ecosystems frequently display the coexistence of diverse species under resource competition, typically resulting in skewed distributions of rarity and abundance. A potential driver of such coexistence is environmental fluctuations that favor different species over time. How to include and treat such temporal variability in existing consumer-resource models is still an open problem. In this work, we study correlated temporal fluctuations in  species' resource uptake rates -- i.e. \textit{metabolic strategies} -- within a stochastic consumer-resource framework. In a biologically relevant regime, we are able to find analytically the species abundance distributions through the path integral formalism. Our results reveal that stochastic dynamic metabolic strategies induce community structures that align more closely with empirical ecological observations. Within this framework, ecological communities show a higher diversity than expected under static competitive scenarios. We find that all species become extinct when the ratio of the number of species to the number of resources exceeds a critical threshold. Conversely, diversity peaks at intermediate values of the same ratio. Furthermore, when metabolic strategies of different species are different on average, maximal biodiversity is achieved for intermediate values of the amplitude of fluctuations. This work establishes a robust theoretical framework for exploring how temporal dynamics and stochasticity drive biodiversity and community structure.}
\end{abstract}

\maketitle

\section{Introduction}

Highly biodiverse ecosystems are ubiquitous~\cite{pomeroy1988concepts, Wilson1992}. Examples include \resub{communities of} plants, birds, animals, and microorganisms. The astonishing richness of biodiversity is reflected in the fact that around 1.75 million species have been identified; however, this figure represents only about 20\% of the estimated total that may exist~\cite{May1988, Wilson1992, niche-neutral-1}. Within single trophic levels in the food web, intricate interactions leads to competition 
among numerous species~\cite{McCann2020, arditi2012species}. 

The {\it competitive exclusion principle} (CEP)~\cite{gause1934, hardin1960} asserts that species competing for the same limiting resource cannot stably coexist \resub{at} equilibrium: one species invariably excludes all others~\cite{Levin1970}. Mathematically, \resub{the CEP}
is supported by the consumer-resource (CR) model~\cite{macarthur1970} describing the population dynamics of species competing for 
resources~\cite{rosenzweig1963, macarthur1970, tilman1982,chase2003}. In stark contrast, Hutchinson’s ``paradox of the plankton'' defies this prediction, with thousands of plankton species coexisting despite feeding on a common pool of limited resources~\cite{hutchinson1961, Plankton-1, Plankton-2}. Similar observations are evident in microbial ecosystems~\cite{soil,seppi2023emergent,pasqualini2024emergent} and terrestrial plant communities~\cite{trees, tilman1982}. On the other hand, the empirical data reveal that  typically most individuals in ecological communities belong to a few species~\cite{Volkov2007,Callaghan2023}, i.e. most of the species composing an ecosystem are rare ~\cite{hubbell2013tropical, van2024}. It is thus of paramount interest to uncover the physical mechanisms that promote and sustain such biodiverse ecosystems,  while also leading to realistic patterns of community structure~\cite{Chesson_mechanism,Grilli2020}.

The coexistence of numerous species despite limited resources is often attributed to specific mechanisms such as resource partitioning~\cite{RP-1,RP-2}, cross-feeding \resub{in microbial communities}~\cite{pfeiffer2004,cross-feeding-2}, predation \resub{in food webs}~\cite{Paine}, and chaotic population dynamics \resub{in plankton communities}~\cite{Chaos}. In recent years, the classical CR model has been extended to incorporate additional mechanisms, including metabolic trade-offs~\cite{posfai2017}, spatial coarse graining~\cite{Gupta_CEP, Gore}, dynamic metabolic adaptation~\cite{leonardo}, and Liebig's law of the minimum~\cite{Chaos} which states that species growth is determined by the scarcest available resource rather than the total resource pool.

While all these mechanisms are realistic and probably important, \resub{we still lack a comprehensive theoretical framework which takes into account the overall influence 
of temporal variability on species' abilities to uptake resources.} In fact, ecosystems are generally influenced by fluctuations of various origins~\cite{Lande}. For example, vital rates of organisms may change significantly over time, due to adaptation dynamics~\cite{shimada2010,raubenheimer2012}, environmental variations~\cite{Giroud2020,pinek2020, Bernhardt2020}, and eco-evolutionary processes~\cite{hairston2005,carroll2007,fronhofer2023}. These temporal variations do not allow population dynamics to reach equilibrium~\cite{morozov2020} and create conditions under which species can coexist by exploiting different niches or by experiencing fluctuating selection pressures that prevent any one species from dominating indefinitely~\cite{macarthur1958, connell1978}.  
Hutchinson~\cite{hutchinson1961} notably proposed that the interplay between environmental and biological time scales could prevent competitive exclusion, suggesting that equilibrium would be avoided, and the CEP violated, when these two time scales are relatively close compared to others. \resub{In these cases, species richness and thus biodiversity would be increased with respect to other scenarios in which time scales are well-separated.} This idea \resub{broadly} aligns with the \textit{intermediate disturbance hypothesis} (IDH)~\cite{connell1978}, which posits that biodiversity is maximized at intermediate levels of disturbance. \resub{One source of disturbances is found in} fluctuations induced by environmental changes or shifts in resource availability, \resub{which can} disrupt competitive hierarchies, preventing dominant species from excluding others and thereby fostering coexistence~\cite{klausmeier2010successional}.
This hypothesis provides a framework for understanding how biodiversity is maintained across a variety of ecosystems and emphasizing the dynamic interplay between competition, resource availability, and environmental stability. Despite these insights, for ecological communities composed of numerous species and resources, a comprehensive understanding of the effects of stochastic fluctuations on species abundance distributions predicted by the CR model is still lacking. Moreover,
the theoretical validity of the IDH \resub{and related patterns of biodiversity} remain subject of active discussion~\cite{fox2013intermediate,sheil2013defining,fox2013intermediate_reply,mancuso2021environmental}.

To address these gaps, we introduce stochastic fluctuations in species' metabolic strategies \resub{(i.e., their resource uptake rate\footnote{\resub{We note that, throughout the literature, these uptake rates are also referred to by other terms, such as \textit{feeding strategies}, \textit{allocation strategies} or \textit{feeding preferences}. In this work, we keep to the term metabolic strategies, used also in Refs.~\cite{posfai2017,PaccianiMori2020,Gupta_CEP,caetano2021evolution,BatistaToms2021}.}})}, representing random variations of the expressed metabolic functions.
Recent studies have explored random Lotka-Volterra~\cite{Azaele2024generalized,suweis2024generalized,SLVM-1,Galla2018,ferraro2025synchronization} and random CR~\cite{SCRM-1, BatistaToms2021} models to examine the effective population densities of large ecological communities. However, in CR models, only time-independent (quenched) stochastic \resub{metabolic strategies have been considered~\cite{tikhonov2017collective,cui2020effect,BatistaToms2021}}. In our framework, by contrast, metabolic strategies are modeled as a \resub{Gaussian} random process with temporal fluctuations characterized by exponential autocorrelation. \resub{Here, we consider all possible sources of stochasticity as included into the random dynamics of metabolic strategies. This might happen, for example, when a fluctuating essential resource is integrated out of the model, thus leaving a statistical signature in the rate of consumption of all other resources~\cite{stomp2008timescale}. Throughout this work, we will broadly refer to such stochastic factors as of \textit{environmental fluctuations}. Our approach is complementary to that of recent works~\cite{bloxham2024biodiversity,wang2024fitness}, where environmental fluctuations are described through time-dependent resource supply.}

Using the path integral formalism~\cite{ferraro2025exact}, we derive an {\it effective consumer-resource dynamics} in the limit of infinitely many species and resources. These equations enable us to compute the species abundance distribution (SAD) for general correlation times with minimal approximations, and exactly for vanishing correlation times. Our analysis\resub{, which employs community evenness as an indicator of effective biodiversity,} reveals that:
(i) predicted species distributions closely align with empirical observations~\cite{Grilli2020}, showing Gamma-like distributions for short correlation times \resub{of fluctuating metabolic strategies};
(ii) the CEP is violated for finite correlation times, with stronger violations for shorter correlation times;
(iii) maximal CEP violation occurs when competition strength is intermediate, leading to an \textit{intermediate competition hypothesis};
(iv) in the presence of non-neutral effects, optimal biodiversity is achieved at intermediate levels of fluctuation amplitude, a result related to -- but independent from -- the \textit{intermediate disturbance hypothesis}~\cite{connell1978}.

\section{Methods}
Our aim is to study the effects of temporal fluctuations of organisms' feeding rates, i.e., of \textit{metabolic strategies}, on the patterns and coexistence of the community. 
We describe \resub{two trophic levels}, meaning \resub{one level comprising} organisms which interact primarily through the utilization of \resub{another level constituted by a} common pool of nutrients, e.g.,  \resub{herbivores feeding on primary producers or bacteriophages feeding on bacteria}. More explicitly, we consider a consumer-resource model of $S$ species competing for $R$ biotic and substitutable resources. 
At time $t$, the population size of species $\sigma = 1,2,\dots,S$ is $n_\sigma(t)$, and the concentration of resource $i = 1,2,\dots,R$ is $c_i(t)$.

These quantities evolve according to the following equations:
\begin{subequations}\label{eq:model_full}
\begin{align}
\dot n_\sigma &= n_\sigma\bigg[\sum_{i=1}^R\alpha_{\sigma,i}(t) c_i - \delta_\sigma\bigg] + \lambda_n\ ,\\\
\dot c_i &= \mu_i c_i\bigg(1 - \dfrac{c_i}{\kappa_i}\bigg) - c_i\sum_{\sigma=1}^S n_\sigma \alpha_{\sigma,i}(t) + \lambda_c\ , \label{eq:color0}\\
\alpha_{\sigma,i}(t) &= \dfrac{\bar{\alpha}}{S} + \dfrac{\Delta_{\sigma,i} + \Sigma Z_{\sigma,i}(t)}{\sqrt{S}}\ ,  \label{eq:color1}\\
\dot Z_{\sigma,i} &= - \dfrac{Z_{\sigma,i}}{\tau} + \dfrac{\sqrt{1+2\tau}}{\tau}\xi_{\sigma,i}(t)\ ,
\label{eq:color2}
\end{align}
\end{subequations}
where \resub{$\{\alpha_{\sigma,i}(t)\}_{i=1}^R$ is the metabolic strategy characterizing the rate of consumption of each} resource $i$ by species $\sigma$ at time $t$, $\delta_\sigma$ is the death rate
of species $\sigma$, $\mu_i$ and $\kappa_i$, respectively, are the growth rate and carrying capacity of $i$-th resource. Further, both consumers and resources can immigrate in the system with constant rates $\lambda_n$ and $\lambda_c$, respectively, which should be set to zero for isolated systems.

\resub{The dynamical metabolic strategies~\eqref{eq:color1} incorporate different contributions:} all species share the same baseline level for metabolic strategies, given by $\bar\alpha/S$; species-specific variations from that baseline are modeled through $\Delta_{\sigma,i}/\sqrt{S}$, which are
quenched independent and identically distributed normal random variables, $\Delta_{\sigma,i}\sim \mathcal{N}\big(0,\Delta^2\big)$.
The scaling with $S$ in Eq.~\eqref{eq:color1} is necessary so that the limit $S \to +\infty$ is well-defined~\cite{SLVM-1,Galla2018}.
In Eqs.~\eqref{eq:color1} and \eqref{eq:color2},  $\xi_{\sigma,i}(t)$ are independent Gaussian white noises with zero mean and delta-correlated in time: $\langle \xi_{\sigma,i}(t) \xi_{\sigma',i'}(t') \rangle = 2\delta_{\sigma,\sigma'} \delta_{i,i'}\delta(t-t')$, such that each \resub{stochastic process} $Z_{\sigma,i}$ at stationarity behaves as a colored-noise with zero mean and exponentially-correlated in time:
\begin{align}
   \langle Z_{\sigma,i}(t)~Z_{\sigma',i'}(t')\rangle &= \delta_{\sigma,\sigma'} \delta_{i,i'}\dfrac{1 + 2\tau}{2\tau} e^{-|t-t'|/\tau} \\&\equiv \delta_{\sigma,\sigma'} \delta_{i,i'} q(|t-t'|)\ , \label{z-corr}
\end{align}
\resub{where $\tau$ defines the correlation time of the metabolic strategies, $\alpha_{\sigma,i}$, through $Z_{\sigma,i}$.}
 \resub{In Eq.~\eqref{eq:color1}, the parameter $\Sigma$ regulates the amplitude of the colored-noise fluctuations.}

In the limits of vanishing and infinite correlation time $\tau$, the temporal correlation in Eq.~\eqref{z-corr} behaves as
\begin{align}\label{eq:OU_scaling}
   \langle Z_{\sigma,i}(t)~Z_{\sigma',i'}(t')\rangle \to \delta_{\sigma,\sigma'} \delta_{i,i'}
\begin{cases}
\delta(t-t')&~{\rm as}~\tau\to 0\\
1&~{\rm as}~\tau\to \infty
\end{cases}\ .
\end{align}
\resub{Therefore, in the limit of vanishing correlation time (i.e., $\tau\to 0$), each $\alpha_{\sigma,i}$ behaves as a Gaussian white noise [see Eq.~\eqref{eq:color1} and \eqref{eq:OU_scaling}]:
\begin{align}
   \alpha_{\sigma,i}(t) \to \dfrac{\bar{\alpha}}{S} + \dfrac{\Delta_{\sigma,i} + \Sigma \xi_{\sigma,i}(t)}{\sqrt{S}}\ ,
\end{align}
whereas in the limit of very large correlation time (i.e., $\tau\to \infty$) the metabolic strategies become independent and identically distributed Gaussian random variables that are fixed in time (quenched) with average $\bar\alpha/S$ and variance $(\Delta^2+ \Sigma^2)/S$ [see Eq.~\eqref{eq:color1} and \eqref{eq:OU_scaling}]. Therefore, each $ \alpha_{\sigma,i}(t)$ fluctuates about $\bar\alpha/S + \Delta_{\sigma, i}/\sqrt{S}$ on time scales longer than $\tau$, whereas on shorter time scale it behaves like a quenched Gaussian random variable.  We further note from Eq.~\eqref{eq:color1} that in the absence of quenched stochasticity of the metabolic strategies, i.e. $\Delta_{\sigma,i} = 0 \ \forall \sigma,i$,  all $S$ species are equivalent on average over time scales $T \gg \tau$. However, over time scales $T \ll \tau$, interspecific differences are significant, hence the model is not {\it neutral} in a strict sense. 

We remark that modeling metabolic strategies' fluctuations as Gaussian colored noise is a minimal choice which enable us to describe stochastic effects resulting various additive contributions arising from many diverse sources of environmental variability.}
\resub{Further, the prefactor in front of the Gaussian noise $\xi_{\sigma,i}$ in Eq.~\eqref{eq:color2} is the simplest choice that leads to the limiting behavior in Eq.~\eqref{eq:OU_scaling}, which we are interested in. We stress that, nonetheless, different choices consistent with the limiting behavior~\eqref{eq:OU_scaling} would yield qualitatively similar results.}

\begin{figure}
    \centering
    \includegraphics[width=1\linewidth]{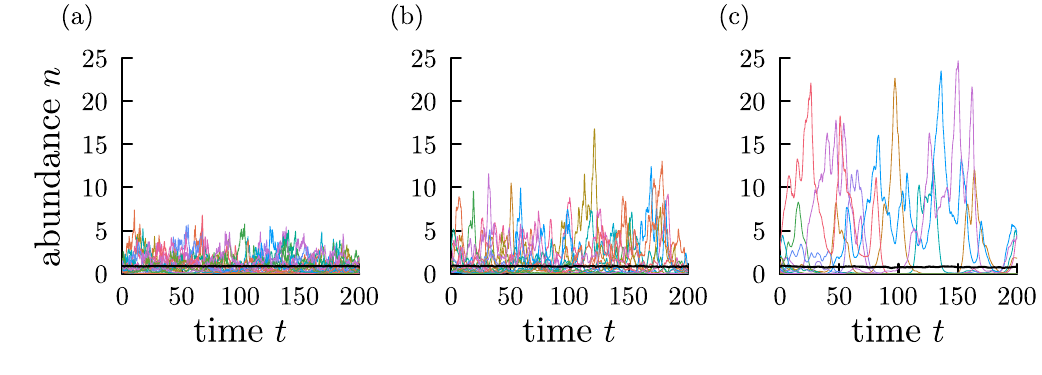}
    \caption{Examples of trajectories of $\{n_\sigma\}$, solutions of Eqs.~\eqref{eq:model_full}, for $\tau=0.1$ (a), $\tau=1$ (b) and $\tau=10$ (c), with all remaining parameters being equal (the values of all other parameters are given in appendix~\ref{si:numericals}); in particular, $\Delta = 0$\resub{, $S=125$ and $R=50$}. Colored lines show abundances of individual species, of which only 20 are shown for each instance, while the thick black lines represent the average species abundance.}
    \label{fig:trajectories_examples}
\end{figure}

We show some examples of solutions of Eq.~\eqref{eq:model_full} in Fig.~\ref{fig:trajectories_examples}, displaying the qualitative differences in trajectories for different values of $\tau$: fluctuations in species abundances become larger, and correspondingly turnover between species slows down.

In what follows, for convenience, we set $\delta_\sigma = \delta$, $\mu_i = \mu$, and $\kappa_i = \kappa$ for all consumers and resources. We regard all quantities as dimensionless. We note that choosing to scale metabolic strategies with $R$ instead of $S$ would produce identical results up to a redefinition of constants: $\bar\alpha \to \bar\alpha S/R,~ \Delta \to \Delta \sqrt{S/R},~\Sigma \to \Sigma \sqrt{S/R}\ .$

In the following sections, we derive the effective dynamics for Eq.~\eqref{eq:model_full} in the limit of infinite species and resource types, with a finite \resub{\textit{species-resources ratio} $\nu\equiv S/R$. Given that the total available resource per species decreases both with fixed $R$ and increasing $S$, or with fixed $S$ and decreasing $R$, we henceforth also refer to $\nu$ as to the \textit{competition strength} of the ecosystem}. Then, we give the expression for the stationary distribution of abundances of both species and resources  \resub{in the absence of quenched disorder}; 
these results allow us to illustrate some key predictions of this model.
In particular, in the limit \resub{of vanishing correlation time (i.e., $\tau \to 0$)}, we can write down the exact stationary distributions of the process, showing directly their dependence on the model's parameters. 
For $\tau > 0$, we obtain approximated stationary distributions, which depend on the statistics of the full system. Later, we explore the case of both quenched and annealed disorders, illustrating the effects of their interplay.

\subsection{Effective equations in the thermodynamic limit}
\label{eff-eqn}
Equation~\eqref{eq:model_full} describes the dynamics of the population of $S$ species competing for $R$ resources in the presence of stochastic metabolic strategies $\alpha_{\sigma,i}(t)$.
While an exact solution of the full model, Eq.~\eqref{eq:model_full}, is not achievable, following Refs.~\cite{suweis2024generalized,BatistaToms2021} we utilize {\it dynamical mean field theory} (DMFT) to derive {\it effective dynamical equations} that in a biologically relevant regime we will able to solve analytically. In the limit of infinite numbers of species $S$ and resources $R$, with their ratio $\nu=S/R$ being fixed and finite, these stochastic differential equations describe the evolutions of the density, $n$, of a representative species and of the concentration of a representative resource, $c$.  
Here, each differential equation
is driven by a self-consistent noise, where the latter represents the dynamical effect of all other (infinitely many) species and resources. 
Here, we report the DMFT equations, while the details of their calculations are relegated to Appendix~\ref{si:DMFT_derivation}. 
The effective equations read
\begin{widetext}
\begin{subequations}\label{eq:DMFT_effective_dynamics}
\begin{align}
\dot{n} &= n \left[ \frac{\bar\alpha}{\nu}\langle c \rangle - \delta -\frac{1}{\sqrt{\nu}} \xi_{n}\right] + n\left[ \frac{1}{\nu}\int_0^t~ dt' \Big(\Delta^2 + \Sigma^2 q(|t-t'|)\Big) G_c(t,t') n(t') + h_n \right] + \lambda_n\ , \label{eff:n-eqn} \\
\dot{c} &= \mu c\left(1-\frac{c}{\kappa}\right) - c\left[\bar\alpha\langle n \rangle + \xi_c + \int_0^t~ dt'\left(\Delta^2 + \Sigma^2 q(|t-t'|)\right)G_n(t,t')c(t') - h_c \right] + \lambda_c\ ,\label{eff:c-eqn} 
\end{align}
\end{subequations}
\end{widetext}
where the noises \finalresub{$\xi_n$ and $\xi_c$} satisfy the following self-consistent equations:
\begin{subequations}
\label{eqn:8}
    \begin{align}
        \langle \xi_n \rangle &= \langle \xi_c \rangle = 0\ , \\
\langle \xi_n(t) \xi_n(t') \rangle &=\Big(\Delta^2 + \Sigma^2 q(|t-t'|)\Big) \langle c(t)c(t') \rangle\ , \\
\quad \langle \xi_c(t) \xi_c(t') \rangle &= \Big(\Delta^2 + \Sigma^2 q(|t-t'|)\Big) \langle n(t)n(t') \rangle\ ,
    \end{align}
\end{subequations}

with the two-points time correlation function of the driving noise, $q(|t-t'|)$, defined in Eq.~\eqref{z-corr}.
Averages with angular brackets in Eq.~\eqref{eqn:8}, are understood to be taken with respect to realizations of the effective dynamics~\eqref{eq:DMFT_effective_dynamics}. The auxiliary fields, $h_n(t)$ and $h_c(t)$, have been introduced for mathematical convenience, as they allow us to compute the response functions:

\begin{subequations}
\begin{align}\label{eq:DMFT_response_function}
    G_n(t,t') = \Bigg\langle \frac{\delta n(t)}{\delta h_n(t')} \Bigg\rangle \ , \\
    G_c(t,t') = \Bigg\langle \frac{\delta c(t)}{\delta h_c(t')} \Bigg\rangle \ .
\end{align}
\end{subequations}

Notice that Eqs.~\eqref{eff:n-eqn} and \eqref{eff:c-eqn} are only coupled through the self-consistent conditions~\eqref{eqn:8}, which makes the effective equations~\eqref{eq:DMFT_effective_dynamics} amenable to analytic calculations. \resub{Hereafter, we set $h_n(t) = h_c(t) \equiv 0$.}

In the following, we investigate this effective dynamics at steady-state, giving an analytical description in the purely annealed case, i.e.,  $\Delta = 0$, unless stated otherwise.
Disordered consumer-resources models have been studied with quenched disorder present in both metabolic strategies and resource carrying capacities \cite{BatistaToms2021}; in the present work, we focus on fluctuating metabolic strategies, investigating the role of temporal fluctuations with general autocorrelation time.  Our results recover those in Ref.~\cite{BatistaToms2021} when taking the limit $\tau \to +\infty$.

\subsection{Exact results in the white noise limit}
\label{sec:wnl}
In this section, we solve the Eqs.~\eqref{eff:n-eqn} and \eqref{eff:c-eqn} to obtain the species abundance distribution in the white noise limit: $\tau\to 0$. Specifically, we set the auxiliary fields $h_{n,c}$ equal to 0 and take the limit $\tau \to 0$ [i.e., $q(|t-t'|)\to \delta(t-t')$], and we obtain the exact steady-state distributions of $n$ and $c$ (see Appendix~\ref{si:WNL} for more details):
\begin{subequations}
\label{eq:wnl_SAD}
    \begin{align}
p_{\rm st}^{(n)}(n;\tau = 0) &\propto n^{-1+d_n} e^{-n/b_n -\ell_n/n}\ , \\
p_{\rm st}^{(c)}(c;\tau = 0) &\propto c^{-1+d_c} e^{-c/b_c -\ell_c/c}\ ,
\end{align}
\end{subequations} 
where all the constants are defined in Appendix~\ref{si:WNL}.
Setting the immigration rates $\lambda_{n} = \lambda_{c}  = 0$ -- signifying the system is closed -- results in $\ell_n = \ell_c = 0$.
In this case, the population and resource concentration distributions~\eqref{eq:wnl_SAD} become Gamma distributions, whereas one obtains \resub{truncated} Gaussian distributions for the quenched disorder system which corresponds to the opposite limit $\tau\to \infty$ and has been studied in Ref.~\cite{BatistaToms2021}. Further, this allows us to obtain closed-form expressions for the first two moments of $n$ and $c$. Solving these moment equations, we can express $\langle n\rangle$ and $\langle c\rangle$ in terms of the models' parameters, yielding explicit analytical predictions for all steady-state statistics. For $\langle n \rangle$ to be positive (or equivalently for $d_n$ to be positive, see Appendix~\ref{si:WNL}), the ratio of number of species to the resources has to be smaller than a critical $\nu_c$ , i.e., $\nu < \bar \alpha \kappa /\delta \equiv \nu_c$, above which all species become extinct.

\subsection{Approximate stationary distributions for general correlation time, $\tau$}
\label{sec:UCNA}

Generally it is difficult to exactly calculate the stationary-state 
distributions of processes driven by colored noise. However, for one-dimensional process driven by Gaussian noise with exponential correlation (i.e., an Ornstein-Uhlenbeck process), the uniform colored-noise approximation (UCNA)~\cite{Jung1987} provides a way to compute the stationary-state distribution. Indeed, the effective variables $n$ and $c$ in Eq.~\eqref{eq:DMFT_effective_dynamics} are only coupled through self-consistent conditions~\eqref{eqn:8}, so that the effective dynamics comprises two one-dimensional equations. Furthermore, the correlation functions of both $n$ and $c$ have approximately exponential form, with characteristic times $\tau_n$ and $\tau_c$ respectively.  We apply the UCNA and obtain the approximated stationary distributions of $n$ and $c$ (see Appendix~\ref{si:UCNA} for calculations):
\begin{widetext}
\begin{subequations}\label{eq:stat_distr_UCNA}
\begin{align}
p_{\rm st}^{(n)}(n;\tau) &\propto  \big(n + A_n\big) n^{-1 + \frac{2\tau}{1+2\tau} B_n} 
\exp\bigg[\frac{2\tau}{1+2\tau} C_n \bigg(-\frac{1}{2} n^2 + D_n n\bigg)\bigg]\ , \\
p_{\rm st}^{(c)}(c;\tau) &\propto \big(c + A_c\big) c^{-1 + \frac{2\tau}{1+2\tau} B_c} 
\exp\bigg[\frac{2\tau}{1+2\tau} C_c \bigg(-\frac{1}{2} c^2 + D_c c\bigg)\bigg]\ .
\end{align}
\end{subequations}
\end{widetext}
Both distributions in Eq.~\eqref{eq:stat_distr_UCNA} interpolate between a Gamma and a \resub{truncated} Gaussian distribution. The various parameters appearing in these distributions, $(A_x,B_x,C_x,D_x)_{x\in\{n,c\}}$, depend on the first and second moments of $n$ and $c$, as well as their correlation times and all of the models' parameters, and are reported in Table~\ref{tab:ucna_constants} in Appendix~\ref{si:UCNA}.
Additionally, the UCNA leads to the definition of new constants, on which we comment here briefly. 
Two of these constants, $\chi_n$ and $\chi_c$, are related to the memory terms of the effective dynamics [i.e., the integrals in Eq.~\eqref{eq:DMFT_effective_dynamics}] through the following simplifying assumption:
\begin{equation}
    \int_0^t G_x(t,t')q(t,t')y(t')dt' \approx \chi_x y(t) \ ,
\end{equation}
for  $(x,y) \in \big\{(n,c),(c,n)\big\}$. Moreover, the parameters of the distributions in Eq.~\eqref{eq:stat_distr_UCNA} depend on two emerging time scales, given by $\bar\tau_x \equiv (\tau^{-1}+\tau_x'^{-1})^{-1}$, which is dominated by the smallest between $\tau$ and $ \tau'_x =  \tau_x \langle x^2 \rangle/(\langle x^2 \rangle - \langle x \rangle^2)$, for $x = n,c$. The origin of these time scales is to be found in the assumption that autocorrelation functions of $n$ and $c$ are exponential at stationarity, as reported in Appendix~\ref{si:UCNA}. While moments and correlation times of $n$ and $c$ can be obtained directly from numerical simulations of Eqs.~\eqref{eq:model_full}, $\chi_n$ and $\chi_c$ are not directly accessible, and better considered as fitting parameters.

\subsection{Evenness as an indicator of effective diversity}
\label{sec:even}

As we are working in the limit of infinite number of species and resources [DMFT, Eqs.~\eqref{eq:DMFT_effective_dynamics}], \resub{it is natural to define fractions of species' number with respect to $S$ (the initial number of species in our model ecosystem).} We thus want to investigate how populations' abundances are spread across species. At stationarity, only $S_{\text{eff}}$ ($\leq S$) species will have non-negligible abundance. If individuals are evenly distributed across all $S$ species then one expects $S_{\text{eff}}\simeq S$; on the contrary, if the distribution is highly uneven, $S_{\text{eff}}\ll S$.

A related challenge in both mathematical models and field observations is defining extinction thresholds. In models like ours, species with low fitness \resub{(i.e. typically negative growth rate)} exhibit exponentially declining abundances in time, never truly reaching extinction but potentially transitioning between rare and abundant states over time~\cite{van2024}. In field observations, finite sample size heavily influences the detection of rare species, making it difficult to determine whether a species is locally \resub{present or} extinct within a specific sampling framework \cite{preston1948,tovo2017upscaling}. To address this, a practical abundance threshold is often applied, but this choice affects observed metrics such as species count. A more robust approach is to weight species abundances based on their magnitudes, avoiding arbitrary thresholds and providing a clearer measure of effective diversity.

These considerations naturally lead to the measure of community evenness \cite{mora2016renyi,Zanchetta2024}, \resub{which does not require the introduction of any abundance threshold.} We use evenness to characterize community structure as predicted by our model, Eq.~\eqref{eq:model_full}, in order to analyze the effect of different mechanisms in shaping ecosystems. Evenness describes community proportions, in the sense of relative rarity of species, clarifying the connection between the SAD and relative abundance\resub{s} of an ecosystem ~\cite{mora2016renyi,Hordijk2023,Pigani2024}. 

Given the \resub{state of a community, represented by the} set of all species abundances $\{ n_\sigma\}_{\sigma = 1}^S$, the relative abundance of consumer species $\sigma$ is 
\begin{align}
p_\sigma = \dfrac{n_\sigma}{ \sum_{\sigma'}n_{\sigma'}}\ .
\end{align}
Then, we define the evenness of a community in a given state  $\{ n_\sigma\}_{\sigma}$ as~\cite{Smith1996}%:
\begin{equation}
D = \dfrac{e^{H}}{S}\ ,
\label{eq:evenness}
\end{equation}
where $H$ is the Shannon entropy:
\begin{align}
    H = -\sum_{\sigma=1}^{S} p_\sigma \ln p_\sigma\ .
\end{align}
This definition of evenness is particularly suitable for comparing ecosystems with varying numbers of species, while recognizing that evenness and species richness (\(S\)) represent two distinct yet interconnected dimensions of biodiversity~\cite{Wilsey2000}. The evenness metric satisfies \(0 < D < 1\), where \(D \approx 0\) indicates communities dominated by a few abundant species, and \(D \approx 1\) corresponds to communities where all species have nearly equal abundances\footnote{Since Shannon entropy takes values \(H \in [0, \ln{S}]\), evenness is also bounded, i.e., \(D \in [1/S, 1]\). However, in the limit \(S \to \infty\), \(D\) spans \([0, 1]\).}. \resub{As we show in Appendix~\ref{si:evenness_rarefaction}, evenness as defined in Eq.~\eqref{eq:evenness} directly measures how quickly different species can be detected in a well-mixed community~\cite{chao2015estimating,siegel2004rarefaction}.}

\resub{Overall}, the metric \(D\) can be interpreted as an indicator of the \resub{\textit{effective}} fraction of species, relative to the total \(S\), that make up the bulk of the community. \resub{Moreover,} in the specific context of a consumer-resource model, where the competition strength \resub{(i.e. the species-resource ratio)} is defined as \(\nu = S/R\), the product \(\nu D\) represents the \resub{\it effective} number of species per resource\finalresub{: $\nu D = S_{\text{eff}} / R$}. Therefore, this metric measures the \resub{\textit{effective}} violation of the CEP\finalresub{, which we consider to take place if $\nu D > 1$.}

We have analytically shown that the SAD in the white noise limit is a Gamma distribution [Eq.~\eqref{eq:wnl_SAD}]. In Appendix~\ref{si:evenness} we compute the evenness for this case $n \sim \text{Gamma}\big(d,b\big)$ in the limit of infinite $S$, and obtain $D = e^{-\psi(1+d)}d$, for the the Digamma function $\psi(z) \equiv \frac{d}{dz}\ln(\Gamma(z))$; see Fig.~\ref{fig:evenness_gamma_examples}. This shows that $D$ is a well-defined quantity for highly diverse ecosystems, and \resub{for Gamma distributions} depends only on one parameter of the SAD.

\resub{For the sake of comparison, in Appendix~\ref{si:evenness} we also consider \textit{strict} diversity described by the fraction $\left(S^* / S\right)_{\varepsilon}$, where $S^*$ is the number of species whose abundances lie above the \textit{extinction threshold} $\varepsilon$, and note\footnote{\finalresub{Consider the homogeneous distribution $q = \{q_i\}_i$ and the generic distribution $p=\{p_i\}_i$, both with support on $i = 1,\dots,S^*$; $q_i = 1/S^* \ \forall i \in \{1,\dots,S^*\}$, and $1 \leq S^* \leq S$. The Kullback-Leibler divergence between the two distributions is $\text{KL}(p||q) = \sum_i \left[p_i \ln\left(S^* p_i\right)\right] = -H + \ln\left(S^*\right) \geq 0$, where $H$ is the Shannon entropy of $p$. It follows immediately that $S^* \geq e^H \equiv S_{\text{eff}}$, or equivalently that $\left(S^*/S\right)_{\varepsilon=0} \geq D$.}} that $D \leq \left(S^* / S\right)_{\varepsilon=0}$; as shown numerically (Fig.~\ref{fig:CEP_strict_vs_effective}), this inequality also holds for small but finite $\varepsilon$. }\finalresub{As the CEP in its strict form is violated when $S^*/R > 1$, from the preceding discussion it follows that effective CEP violation always implies strict violation.}

\section{Results}

We analyzed how fluctuating metabolic strategies shape the structure of consumer-resource ecological communities. These metabolic strategies are characterized by several parameters: the heterogeneity among species (\(\Delta\)); the amplitude (\(\Sigma\)) and correlation time (\(\tau\)) of the fluctuations; and the competition strength (\(\nu = S/R\)). \resub{The community structure described by our model is fully captured by the \resub{species abundance distribution (SAD)}. However, in order to describe patterns of species rarity for different realizations of our model, we consider the evenness indicator $D$ as defined in Eq.~\eqref{eq:evenness}. This offers a more nuanced description of biodiversity than presence/absence information (see Fig.~\ref{fig:CEP_strict_vs_effective}). Moreover, we quantify the breaking of the CEP in an effective sense by the metric $\nu D$, which represent the diversity (or effective number of species) supported by each resource.}

\resub{We first considered the case \(\Delta = 0\), where the long-term interspecific differences in metabolic strategies are negligible compared to temporal fluctuations. By solving the full system, we observed that increasing the correlation time \(\tau\) leads to slower but larger fluctuations in species abundances (see Fig.  \ref{fig:trajectories_examples}). As \(\tau \to +\infty\), communities become more uneven, with many rare species and a few abundant ones. This behavior is consistent with the known results for purely quenched disorder~\cite{BatistaToms2021}, where temporal fluctuations are absent.}

\resub{For all numerical results, simulation time $t_{\text{max}}$ has been taken to be much longer than the transient time for most species. Typically, transient times increase  with smaller values of $\Sigma$ and larger values of $\tau$, and for $\Delta >0$.
Thus, accordingly, we increase the numerical simulation time.
Further, in computing stationary quantities we discard data obtained at time $t < f t_{\text{max}} $, where typically $f \gtrapprox 0.5$. Without loss of generality, in all simulations we set $\delta = 1 = \kappa$. All other parameters vary across simulations, including species and resources numbers; however, we used $S,R \gtrapprox 50$ to reduce finite size effects. Further numerical details are reported in Appendix~\ref{si:numericals}.}

\subsection{\(\tau = 0\)}

\begin{figure}
\begin{center}
    \begin{minipage}[t]{0.238\textwidth}
        \centering
        \includegraphics[width=1.0\textwidth]{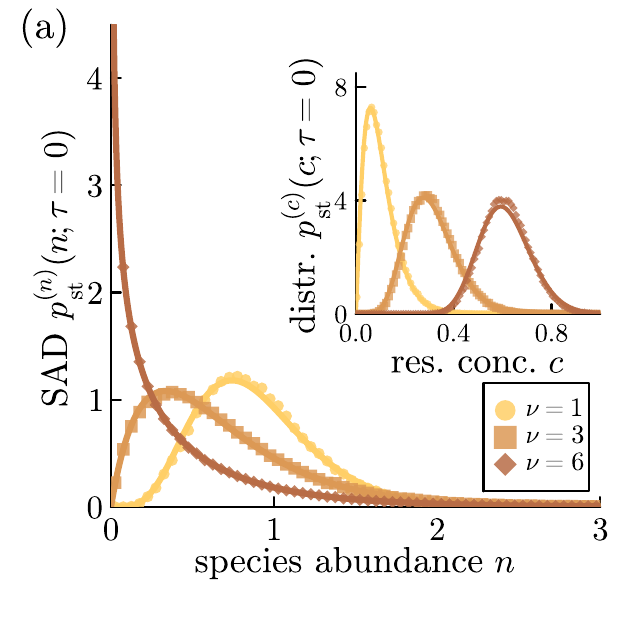}
    \end{minipage}
    \hfill
    \begin{minipage}[t]{0.238\textwidth}
        \centering
        \includegraphics[width=1.0\textwidth]{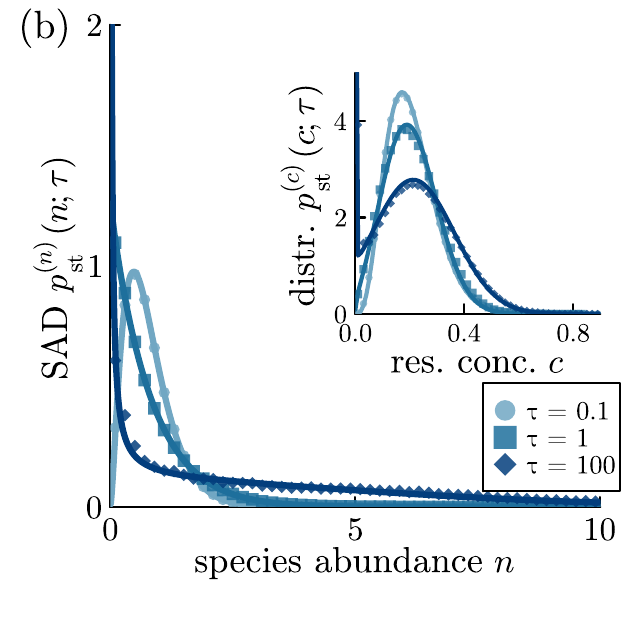}
    \end{minipage}
\end{center}
\caption{(a) Comparison of the empirical stationary distributions \resub{(i.e. species abundance distribution (SAD) and distribution of resource concentration)} obtained from \resub{steady state} simulations of Eq.~\eqref{eq:model_full} (markers) with the corresponding analytical distributions from Eq.~\eqref{eq:wnl_SAD} (solid lines) for different values of \(\nu = S/R\)\resub{, $\nu_c = \bar\alpha \kappa / \delta = 10$, and annealed metabolic strategies ($\Delta = 0$)}. The analytical results are parameter-free predictions of the DMFT. (b) UCNA prediction for stationary distributions of species and resource abundances for \(\nu = 2\) \resub{($\nu_c = 10$ and $\Delta = 0$)} with \(\tau = 0.1\), \(1\), and \(100\).  \resub{Simulation results: Markers. Best-fit of predicted distributions: Solid lines.} Main panels show the distribution \resub{of species abundances} $n$, while the insets show \resub{the distribution of resource concentrations} $c$. All parameters used in the simulations \resub{and further numerical details} are reported in Appendix~\ref{si:numericals}. }
\label{fig:SAD_wnl_ucna}
\end{figure}

We studied analytically the case of vanishing correlation time (\(\tau \to 0\)). We derived a closed-form expression for the species abundance distribution (SAD)~\eqref{eq:wnl_SAD} and an exact functional form for the evenness (Appendix~\ref{si:evenness}). Numerical simulations confirmed the excellent agreement between the theoretical SAD and the  distributions obtained from the model (Fig.~\ref{fig:SAD_wnl_ucna}a). 

For larger values of competition strength \(\nu\), the SAD peak shifts toward \(n=0\), as increasing species \finalresub{number $S$} for a fixed number of resources \finalresub{$R$} leads to lower abundances for most species. Notably, under annealed uncorrelated disorder, the SAD takes the form of a Gamma distribution, consistent with empirical observations from real ecosystems~\cite{azaele2016statistical,Grilli2020, suweis2024generalized}. This result emphasizes the robustness of our model in capturing universal patterns of biodiversity. While our primary focus was on biotic resources, Appendix~\ref{si:abiotic} demonstrates that similar SADs emerge for abiotic resources, further validating the generality of our findings.

\subsection{General \(\tau\)}
For a general $\tau$, a richer phenomenology is observed. Using the Unified Colored Noise Approximation (UCNA), we derived approximate stationary distributions for general \(\tau\) (Eq.~\eqref{eq:stat_distr_UCNA}), which interpolate between Gamma and \resub{truncated} Gaussian shapes. The parameters of the distributions depend on the moments of \(n\) and \(c\) and their correlation times, which we
estimated directly from simulations of the full model, Eq.~\eqref{eq:model_full}. Having no access to the values of $\chi_n$ and $\chi_c$ for generic $\tau$, we used these constants as {\it fitting parameters}. Figure~\ref{fig:SAD_wnl_ucna}b shows that the theoretical (approximated) SAD closely matches numerical results.

 As \(\tau\) increases, the turnover between rare and abundant species slows down (Appendix~\ref{si:local_time}). For this reason, rare species will experience longer periods of decay, leading to even lower abundances.
\finalresub{Concurrently, }\resub{a truncated} Gaussian component with large variance emerges in the SAD, representing abundant species, while rare species form a Gamma-like peak near \(n=0\). \resub{While, for finite $\tau$, these rare species are not extinct,} this result is reminiscent of truncated Gaussian distributions observed in quenched-noise models~\cite{Galla2018, BatistaToms2021}. This is also consistent with the predictions of our model in the limit of long correlation time $\tau \to \infty$.

\subsection{\label{sec:effective_CEP_violation} \resub{Effective} CEP violation}

\begin{figure}[h!]
    \centering
    \begin{minipage}[c]{0.238\textwidth}
        \centering
        \includegraphics[width=\textwidth]{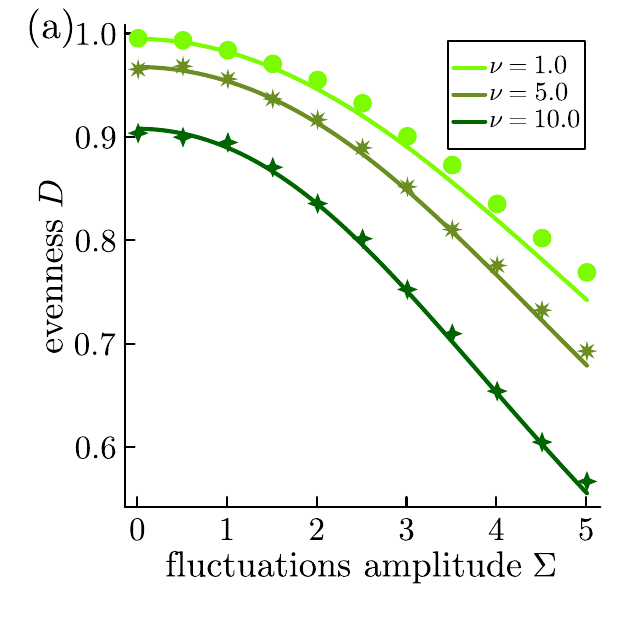}
    \end{minipage}
    \hfill
    \begin{minipage}[c]{0.238\textwidth}
        \centering
        \includegraphics[width=\textwidth]{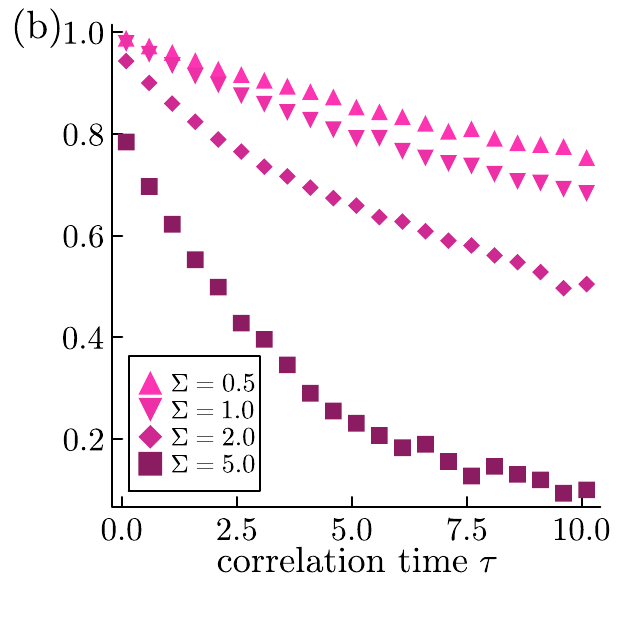}
    \end{minipage}
    \caption{\resub{Dependence of evenness $D$ on the annealed disorder amplitude $\Sigma$ and the correlation time $\tau$
    for purely annealed metabolic strategies ($\Delta = 0$), and $\nu_c = 20$,  
    with white noise (a) and colored noise (b).
    (a) For $\tau \to 0$, and for different values of competition strength $\nu$, we compare the analytical prediction for the evenness of a Gamma distribution, $D=d_n e^{-\psi\left(1+d\right)}$ (solid lines), where $d_n$ is defined in Appendix~\ref{si:WNL}, and results of numerical simulations (markers) obtained as stationary time-averages of evenness~\eqref{eq:evenness} of single realizations of dynamics~\eqref{eq:model_full}. Evenness steeply decrease with fluctuations' size in more competitive ecosystems. (b) For general $\tau > 0$, $\nu = 1$ and different values of sigma $\Sigma$, we similarly compute evenness numerically, showing the interplay of changing fluctuations' size and time scale. All other parameters are reported in Appendix~\ref{si:numericals}.}}
    \label{fig:D_vs_Sigma_tau}
\end{figure}

A key finding of our work is that temporal fluctuations in metabolic strategies allow violations of the CEP. \resub{We use the evenness indicator $D$ (Appendix~\ref{si:evenness}) to describe CEP effective violations.} \resub{For \(\tau \to 0\), the relationship between $\nu$ and $D$ is obtained analytically (Fig.~\ref{fig:D_vs_Sigma_tau}a), while for general $\tau$ it is obtained numerically (Fig.~\ref{fig:D_vs_Sigma_tau}b). In general, for $\Delta = 0$, evenness decreases both with $\Sigma$ and $\tau$, signaling the emergence of the aforementioned ecological pattern of many rare species and few very abundant species.} 

\begin{figure}[h!]
    \centering
    \begin{minipage}[c]{0.238\textwidth}
        \centering
        \includegraphics[width=\textwidth]{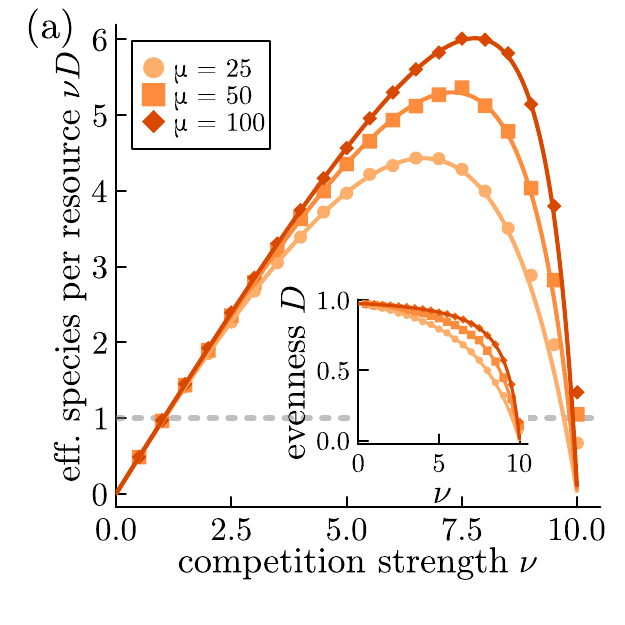}
    \end{minipage}
    \hfill
    \begin{minipage}[c]{0.238\textwidth}
        \centering
        \includegraphics[width=\textwidth]{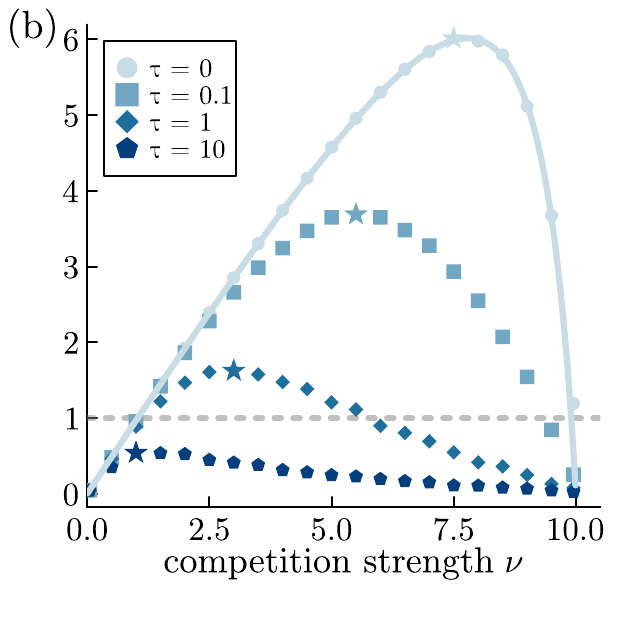}
    \end{minipage}
    \caption{(a) Evenness \(D\) in the white noise limit: \(D\) (inset) and \resub{effective number of species} \(\nu D\) (main panel) as functions of \resub{the competitions strength} \(\nu\) for different values of \(\mu\). Markers: numerical simulations. Solid lines: analytical \resub{evenness $D=d_n e^{-\psi\left(1+d\right)}$, where $d_n$ is defined in Appendix~\ref{si:WNL}}. (b) \(\nu D\) as a function of \(\nu\) for different \(\tau\) values, with $\mu = 100$. Solid line: \resub{exact evenness for \(\tau = 0\), coinciding with the top curve in panel (a)}. Markers: numerical simulations. Stars indicate maximal \(\nu D\) for each \(\tau\) among the values found numerically. CEP is violated when \(\nu D > 1\), a threshold indicated by gray dashed lines. \resub{In both panels, $\Delta = 0$ and $\nu_c = \bar\alpha\kappa / \delta = 10$.} All other parameters used in the simulations are reported in Appendix~\ref{si:numericals}.}
    \label{fig:evenness_vs_nu}
\end{figure}

\resub{As outlined in Section~\ref{sec:even}, the effective violation of the CEP} is quantitatively captured by the metric \(\nu D\) , which represents the effective number of coexisting species per resource. Values of \(\nu D > 1\) indicate that more species \resub{per resource effectively} coexist than allowed under strict competitive exclusion. The \resub{effective} violation of the CEP holds robustly across a wide range of model parameters (Fig.~\ref{fig:evenness_vs_nu}a).

Interestingly, for moderate values of $\tau$, the violation of the CEP is maximized at an intermediate competition strength (Fig.~\ref{fig:evenness_vs_nu}b). In other words, the diversity supported by each resource is higher for intermediate levels of competition, which we refer to as \textit{intermediate competition principle}, a result only emerging for time-dependent interactions. This result complements the \textit{intermediate disturbance principle} by considering endogenous ecological forces, as in our model fluctuations in \resub{metabolic strategies} can lead to effective environmental fluctuations. For $\tau > 0$, competitive exclusion can still be overcome, but as $\tau$ increases the optimal value of $\nu D$ decreases, eventually becoming smaller than $1$, and the optimal $\nu$ moves closer to $1$. This aligns with previous results showing that purely quenched disorder does not break the CEP~\cite{BatistaToms2021}. In all cases, all species go extinct if $\nu \geq \nu_c$; correspondingly, $\nu D \to 0$  as $\nu \to \nu_c$. \resub{This singular regime emerges as a consequence of how the metabolic strategies scale with the number of species in Eq.~\eqref{eq:color1}, therefore we may regard this as a \textit{scaling-induced} CEP.}

\subsection{Combined annealed and quenched disorder}
\begin{figure}
    \centering
    \begin{minipage}[c]{0.238\textwidth}
        \centering
        \includegraphics[width=\textwidth]{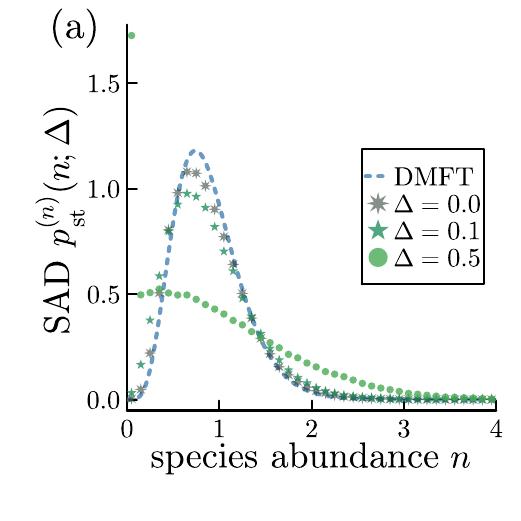}
    \end{minipage}
    \hfill
       \begin{minipage}[c]{0.238\textwidth}
        \centering
        \includegraphics[width=\textwidth]{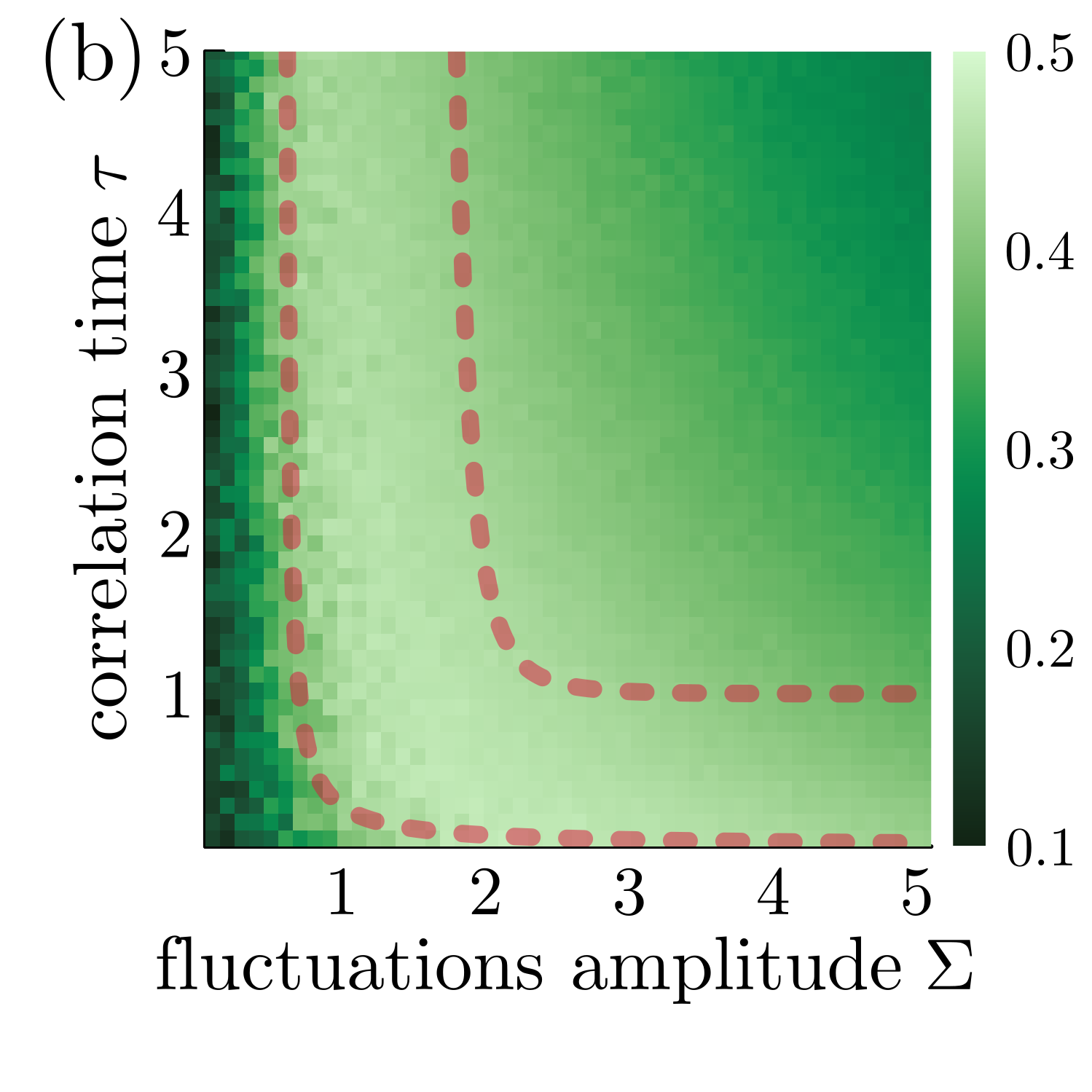}
    \end{minipage}
    \caption{(a) \resub{Species abundance distribution (SAD)} in the case of both quenched and annealed disorder ($\Delta,\Sigma > 0$). Comparison between DMFT results \resub{Eq.~\eqref{eq:wnl_SAD}} in the white noise limit and $\Delta=0$ (dashed line) and \resub{numerical} SADs with $\tau = 0.1$, $\Sigma = \nu = 1$ and different values of $\Delta$ (markers), \resub{with $\nu_c = \bar\alpha \kappa / \delta = 10$}. We highlight the presence of a highly populated bin of low abundances for $\Delta = 0.5$ (marker in the top left corner of the panel). (b) Evenness as function of $\Sigma$ and $\tau$, for $\Delta = 0.1$ \resub{and $\nu = 1$}. Fig.~\ref{fig:phase_diagrams_annealed} in Appendix~\ref{si:local_time} is obtained with identical parameters, save for $\Delta$, which is set to zero. Brighter hues denote more even communities. Each point is obtained by computing the stationary evenness of a single realization of the microscopic dynamics, Eqs.~\eqref{eq:model_full}. Dashed lines in panel (b) contour the region of optimal evenness. All other parameters used in the simulations are reported in Appendix~\ref{si:numericals}.}
    \label{fig:sad_quenched}
\end{figure}

We also investigated the case where both annealed and quenched disorder are present (\(\Delta, \Sigma > 0\)). \resub{Here, given the presence of two types of disorder, the model is not analytically tractable.} Figure~\ref{fig:sad_quenched}a shows how the 
SAD \resub{continuously} transitions from a Gamma distribution (\(\Delta = 0\)) to a truncated Gaussian as \(\Delta \to \Sigma\). \resub{We remark that a similar transition is observed in the purely annealed case as $\tau \to +\infty$. However, we also note that in the present case of combined quenched and annealed disorder this transitions is rather abrupt, since the quenched component becomes dominant (as signaled by the SAD becoming a truncated Gaussian) even when $0 < \Delta < \Sigma$ as $\Delta \to \Sigma$.} Additionally, evenness as a function of \(\Sigma\) and \(\tau\) (Fig.~\ref{fig:sad_quenched}b) reveals a non-trivial region of maximal diversity for \(0 < \Delta \ll \Sigma\), a result analogous to the \textit{intermediate disturbance hypothesis} for the case of endogenous fluctuations. This phenomenon does not occur in the purely annealed case; see Figure~\ref{fig:phase_diagrams_annealed}.

\section{Discussion and Conclusions}

The question of which mechanisms prevents competitive exclusion of species, and how these mechanisms affect ecosystem structure, is a long-standing and important one.
In general, ecosystems are subject to temporal variations in vital rates due to environmental changes and other factors. To capture this dynamic aspect, we studied a consumer-resource model where metabolic strategies are modeled as colored noise, fluctuating around a fixed value with a finite correlation time. This approach describes a phenomenon independent of the specific biological details of organisms, allowing predictions to apply to a broad variety of ecosystems.

Assuming that metabolic strategies of different species share the same baseline value, meaning that long-term differences are negligible compared to temporal fluctuations, \resub{leads to species neutrality over long time scales, while over short time scales, time-dependent interspecific differences  significantly affect community structure and non-neutral effects become important. In this regime, we} derived analytical expressions for the stationary distributions of abundances. In a range of moderate correlation times, the competitive exclusion principle \finalresub{(CEP)} is violated, and the effective number of species per resource is maximized at an intermediate value of competition strength (\(\nu = S/R\)). Further, our model aligns with empirical observations that species abundance distributions (SADs) in most ecosystems follow a robust pattern of many rare species and few common ones, often approximated by Gamma-like distributions \cite{azaele2016statistical,Grilli2020,suweis2024generalized}. Non-zero correlation times introduce a \resub{truncated} Gaussian component to the SAD, which becomes dominant as \(\tau\) approaches infinity, consistent with previous findings~\cite{BatistaToms2021}. While pure Gaussian SADs lack empirical support, moderate correlation times yield distributions where the Gaussian component remains small, resulting in realistic SADs.

Temporal correlations in fluctuations slow the turnover between rare and abundant species, while stronger competition amplifies the impact of each species on others. These combined effects cause SADs to concentrate around rare species, a key characteristic of empirically observed SADs. Our model further predicts that in the white noise limit, all species go extinct when competition strength exceeds a threshold determined by a combination of biological and environmental factors, including baseline metabolic strategies, death rates, and resource carrying capacities. This ecological constraint persists also for finite correlation times.

\resub{It is important to stress that the analytical results demonstrating a violation of the CEP -- both in its strict and effective forms -- are derived under the assumption of vanishing quenched disorder, where species are equivalent on average over long timescales. In this \textit{quasi-neutral} regime, the influence of temporal fluctuations is most clearly disentangled from species-specific heterogeneity. This result implies that the observed CEP violation does not hold for strong interspecific differences. As shown in our numerical results (Fig.~\ref{fig:sad_quenched}a), introducing moderate levels of quenched disorder reduces the extent of CEP violation, signaling that species heterogeneity can counteract the diversity-promoting effects of environmental fluctuations. 
Therefore, long-term non-neutral effects}, such as niche availability and overlap, play a significant role in shaping community structure. In our generalized model, species-dependent baseline metabolic strategies reveal \resub{how} fluctuations of suitable magnitude and correlation time enhance community evenness, reducing the impact of trait differences. These findings are loosely related to the \textit{intermediate disturbance hypothesis} (IDH), but distinct from it, as they encompass temporal variations of diverse origins specifically affecting resource uptake rates. At the same time, our results clarify how different aspects of these variations -- namely, their magnitude and temporal correlation -- interact to define what makes them ``intermediate," an aspect often overlooked in prior work~(see e.g.,  Fig.~1 of Ref.~\cite{connell1978}).
Our analysis highlights a novel perspective: the emergence of maximal diversity at intermediate competition strength. This result demonstrates that an optimal degree of resource competition can foster coexistence by balancing the effects of competitive exclusion and community evenness. These findings offer a quantitative framework to explore the ecological implications of resource competition in a broader context. By linking competition strength and temporal fluctuations, our results open new avenues for understanding how ecosystems maintain biodiversity under varying environmental pressures.

Future studies could extend this framework to include non-substitutable resources~\cite{tamminen2018proteome} or metabolic strategies constrained by realistic trade-offs~\cite{posfai2017,PaccianiMori2020}. Analytical derivations for scenarios combining annealed and quenched disorder would further clarify the conditions under which maximal biodiversity regimes emerge. \resub{At the same time, new techniques need to be developed to make memory terms -- which emerge as unavoidable consequence of correlations between pairwise couplings -- analytically tractable in the stationary regime for general correlation times. Linear models, which are analytically solvable~\cite{ferraro2025exact}, provide an ideal mathematical framework to approach these complex issues.} These developments will help refine our understanding of biodiversity patterns across ecosystems and provide a robust theoretical foundation for ecological management and conservation strategies. \resub{Moreover, experiments probing species' metabolic strategies in highly biodiverse ecosystems would fill an important knowledge gap, thus driving further theoretical developments~\cite{kussell2005phenotypic}.}

\section*{Acknowledgments}

D.Z. gratefully acknowledges MUR and EU-FSE for financial support of the PhD fellowship PON Research and Innovation 2014-2020 (D.M 1061/2021) XXXVII Cycle / Action IV.5 “Tematiche Green”. D.G. gratefully acknowledges the financial support provided by the University of Padova during the academic visit in November 2023, which facilitated this collaboration. D.G. also acknowledges the support from the Alexander von Humboldt foundation. S.S. acknowledges financial support from the MUR - PNC (DD n. 1511 30-09-2022) Project No. PNC0000002, DigitAl lifelong pRevEntion (DARE). A.M. and S.A. acknowledge the support by the Italian Ministry of University and Research (project funded by the European Union–Next Generation EU: PNRR Missione 4 Componente 2, “Dalla ricerca all’impresa,” Investimento 1.4, Progetto CN00000033).

\onecolumngrid

\appendix
\setcounter{secnumdepth}{2}
\section{Derivation of the effective dynamics}\label{si:DMFT_derivation}

In this Appendix we compute the effective equations for the microscopic consumer-resource model, Eqs.~\eqref{eq:model_full}, which we repeat here:
\begin{align}
    \dfrac{\dot n_{\sigma}}{n_{\sigma}} &= \sum_{i=1}^R\alpha_{\sigma,i}(t)~c_i - \delta + h_\sigma\,\label{meq-1}\\
     \dfrac{\dot c_i}{c_i} &= \mu(1-c_i/\kappa)-\sum_{\sigma=1}^S\alpha_{\sigma,i}(t)~n_{\sigma} +h_i,\label{meq-2}\\
     \alpha_{\sigma,i}(t) &= \dfrac{\bar{\alpha}}{S} + \dfrac{\Sigma }{\sqrt{S}}Z_{\sigma,i}(t)\ , \label{alpha-t}\\
    \dot Z_{\sigma,i} &= -\dfrac{Z_{\sigma,i}}{\tau} + \dfrac{\sqrt{1+2\tau}}{\tau} \xi_{\sigma,i}(t)\ .~\label{Z-t}\\
\end{align}
 Here $\sigma\in\{1,2,\dots,S\}$, $i\in\{1,2,\dots,R\}$. To make calculations less cumbersome, we neglect immigration terms; this is a matter of convenience, and they can be safely reinserted at the end of the derivation. Similarly, in this derivation, we omit the quenched disorder. In fact, the annealed and quenched disorders are integrated out independently, but give rise to analogous terms; thus, the effective equations~\eqref{eq:DMFT_effective_dynamics} are recovered from the end results of this calculation simply by performing the substitution $\Sigma^2 q(|t-t'|) \to \Delta^2 + \Sigma^2q(|t-t'|)$. Finally, the auxiliary fields $h_\sigma$ and $h_i$ have been introduced to later compute response functions.

We will calculate the dynamical moment generating function as 
\begin{align}\label{eq:MGF_definition}
    \mathcal{Z}[\psi,\phi]\equiv \overline{\bigg\langle e^{i\sum_\sigma \int~dt~\psi_\sigma(t)n_\sigma(t)+i\sum_j \int~dt~\phi_j(t)c_j(t)} \bigg\rangle},
\end{align}
where the angular brackets indicate the average over paths which are solutions of Eqs.~\eqref{meq-1} and \eqref{meq-2} for a given realization of $\alpha_{\sigma,i}(t)$, and the overhead bar stands for the average over ensemble of realizations of $\alpha_{\sigma,i}(t)$; $\mathcal{Z}[0,0]=1$ by normalization of the path's probability. $\alpha_{\sigma,i}(t)$, whose dynamics is given by Eqs.~\eqref{alpha-t} and \eqref{Z-t}, is distributed at stationarity according to Gaussian distribution with mean $\frac{\bar{\alpha}}{S}$ and variance $\frac{1}{S}\Sigma^2\frac{1+2\tau}{2\tau}$. It also has finite-time exponential correlation, since
\begin{align}
   \langle Z_{\sigma,i}(t) Z_{\sigma',i'}(t')\rangle = \underbrace{\dfrac{1+2\tau}{2\tau} e^{-|t-t'|/\tau}}_{q(|t-t'|)}\delta_{\sigma,\sigma'}\delta_{i,i'}\ . \label{z-corr-t}
\end{align}

The average over paths in Eq.~\eqref{eq:MGF_definition} is written as
\begin{align}
    \langle (\cdots) \rangle &= \int \mathcal{D}[n]\mathcal{D}[c]~e^{i\sum_\sigma \int~dt~\psi_\sigma(t)n_\sigma(t)+i\sum_j \int~dt~\phi_j(t)c_j(t)}\prod_{\sigma,t}\delta\bigg(\dfrac{\dot n_{\sigma}(t)}{n_{\sigma}(t)}- \sum_{j}\alpha_{\sigma,j}(t)c_j(t) + \delta-h_\sigma(t)\bigg) \nonumber\\
    &\times\prod_{j,t}\delta\bigg(\dfrac{\dot c_j(t)}{c_j(t)}- \mu[1-c_j(t)/\kappa]+\sum_{\sigma}\alpha_{\sigma,j}(t)n_{\sigma}(t)-h_j(t)\bigg), \label{z-path}
\end{align}
where the $\prod_{\sigma,t}$ indicates the product of over all $\sigma$ indices and over all times $t$. We use the integral representation of the Dirac Delta function,
\begin{align}
    \delta(y-y_0) \propto \int e^{i\hat{y}(y-y_0)} d\hat{y} \ ,
\end{align}
in Eq.~\eqref{z-path}, leading to the following rewriting:
\begin{align}
\langle (\cdots) \rangle &= \int \mathcal{D}[n,\hat{n}]\mathcal{D}[c,\hat{c}]~\exp\bigg[i\sum_\sigma \int~dt~\psi_\sigma(t)n_\sigma(t)+i\sum_j \int~dt~\phi_j(t)c_j(t)\bigg]\nonumber\\
&\times\exp\bigg[i\sum_\sigma \int~dt~\hat{n}_\sigma(t)\bigg(\dfrac{\dot n_{\sigma}(t)}{n_{\sigma}(t)}- \sum_{j}\alpha_{\sigma,j}(t)c_j(t) + \delta-h_\sigma(t)\bigg)\bigg] \nonumber\\
&\times\exp\bigg[i\sum_j \int~dt~\hat{c}_j(t)\bigg(\dfrac{\dot c_j(t)}{c_j(t)}- \mu[1-c_j(t)/\kappa]+\sum_{\sigma}\alpha_{\sigma,j}(t)n_{\sigma}(t)-h_j(t)\bigg)\bigg]\\
&= \int \mathcal{D}[n,\hat{n}]\mathcal{D}[c,\hat{c}]~\exp\bigg[i\sum_\sigma \int~dt~\psi_\sigma(t)n_\sigma(t)+i\sum_j \int~dt~\phi_j(t)c_j(t)\bigg]\nonumber\\&\times\exp\bigg[i\sum_\sigma \int~dt~\hat{n}_\sigma(t)\bigg(\dfrac{\dot n_{\sigma}(t)}{n_{\sigma}(t)}+ \delta-h_\sigma(t)\bigg)\bigg] \nonumber\\
&\times\exp\bigg[i\sum_j \int~dt~\hat{c}_j(t)\bigg(\dfrac{\dot c_j(t)}{c_j(t)}- \mu[1-c_j(t)/\kappa]-
h_j(t)\bigg)\bigg]\\
&\times \exp\bigg[-i\sum_\sigma \int~dt~\hat{n}_\sigma(t) \sum_{j}\alpha_{\sigma,j}(t)c_j(t)\bigg]\exp\bigg[i\sum_j \int~dt~\hat{c}_j(t)\sum_{\sigma}\alpha_{\sigma,j}(t)n_{\sigma}(t)\bigg]\ ,
\end{align}
where in the last line we have isolated all terms which contain $\alpha_{\sigma,i}(t)$. From the previous expression, we take the average over realizations of $\alpha_{\sigma,i}(t)$:
\begin{align}
\mathcal{Z}[\psi,\phi]\equiv \overline{\langle (\cdots) \rangle}&= \int \mathcal{D}[n,\hat{n}]\mathcal{D}[c,\hat{c}]~\underbrace{\exp\bigg[i\sum_\sigma \int~dt~\psi_\sigma(t)n_\sigma(t) + i\sum_\sigma \int~dt~\hat{n}_\sigma(t)\bigg(\dfrac{\dot n_{\sigma}(t)}{n_{\sigma}(t)}+ \delta-h_\sigma(t)\bigg)\bigg]}_{A[n,\hat{n}]} \nonumber\\
&\times\underbrace{\exp\bigg[i\sum_j \int~dt~\phi_j(t)c_j(t) + i\sum_j \int~dt~\hat{c}_j(t)\bigg(\dfrac{\dot c_{j}(t)}{c_{j}(t)}- \mu[1-c_j(t)/\kappa]-h_j(t)\bigg)\bigg]}_{B[n,\hat{n}]}\nonumber\\
&\times \underbrace{\overline{\exp\bigg[-i\sum_{\sigma,j}\int~dt~\hat{n}_\sigma(t)\alpha_{\sigma,j}(t)c_{j}(t)\bigg]~~\exp\bigg[i\sum_{\sigma,j} \int~dt~\hat{c}_j(t)\alpha_{\sigma,j}(t)n_{\sigma}(t)\bigg]}}_{\Delta[n,\hat{n},c,\hat{c}]}\ .
\end{align}

Let us rewrite the last term as follows:
\begin{align}
    \Delta[n,\hat{n},c,\hat{c}] &=\overline{\exp\bigg[i\sum_{\sigma,j}\int~dt~[\hat{c}_j(t)n_{\sigma}(t)-\hat{n}_\sigma(t)c_j(t)]\alpha_{\sigma,j}(t)\bigg]}\nonumber\\
     &=\exp\bigg[i\frac{\bar{\alpha}}{S}\sum_{\sigma,j}\int dt [\hat{c}_j(t)n_{\sigma}(t)-\hat{n}_\sigma(t)c_{j}(t)]\bigg] \nonumber\\
    &\times \exp\bigg[-\frac{\Sigma^2}{2S}\sum_{\sigma,j}\int dt \int dt' [\hat{c}_j(t)n_{\sigma}(t)-\hat{n}_\sigma(t)c_{j}(t)][\hat{c}_j(t')n_{\sigma}(t')-\hat{n}_\sigma(t')c_{j}(t')]q(|t-t'|)\bigg]\ , \label{last-line}
\end{align}
where in the last line we used the following result (see  Appendix~\ref{sec:lastline}):
\begin{align}
    \overline{\exp\bigg[\frac{i\Sigma}{\sqrt{S}} \int_0^t dt~A(t)Z(t)\bigg]}  = \exp\bigg[-\dfrac{\Sigma^2}{2S} \int_0^t dt_1\int_0^t dt_2 ~A(t_1)A(t_2)q(|t_1-t_2|)\bigg]\ . \label{pf-lastline}
\end{align}

Thus, 
\begin{align}\label{eq:Delta_path_integral}
    \Delta[n,\hat{n},r,\hat{c}]&=\exp\bigg[i \bar{\alpha}S \int~dt~[\rho_{n}(t)\lambda_c(t)-\rho_c(t)\lambda_n(t)]\bigg]\nonumber\\
    &\times\exp\bigg[-\frac{\Sigma^2 S}{2}\int~dt~\int~dt'\big[Q_n(t,t')L_c(t,t')+Q_c(t,t')L_n(t,t')-2K_n(t,t')K_c(t,t')\big] q(|t-t'|) \bigg] \ .
\end{align}

In Eq.~\eqref{eq:Delta_path_integral} we defined the following order parameters:
\begin{align}
    \rho_n(t) = \dfrac{1}{S} \sum_\sigma n_\sigma(t) &\qquad\qquad \lambda_n(t) = \dfrac{1}{S} \sum_\sigma \hat{n}_\sigma(t)\label{cond-1}\\
    \rho_c(t) = \dfrac{1}{S} \sum_i c_i(t) &\qquad\qquad \lambda_c(t) = \dfrac{1}{S} \sum_i \hat{c}_i(t)\\
    Q_n(t,t') = \dfrac{1}{S} \sum_\sigma n_\sigma(t)n_\sigma(t') &\qquad\qquad L_n(t,t') = \dfrac{1}{S} \sum_\sigma \hat{n}_\sigma(t)\hat{n}_\sigma(t')\\
    Q_c(t,t') = \dfrac{1}{S} \sum_i c_i(t)c_i(t') &\qquad\qquad L_c(t,t') = \dfrac{1}{S} \sum_i \hat{c}_i(t)\hat{c}_i(t')\\
    K_n(t,t') = \dfrac{1}{S} \sum_\sigma \hat{n}_\sigma(t)n_\sigma(t') &\qquad\qquad     K_c(t,t') = \dfrac{1}{S} \sum_i \hat{c}_i(t)c_i(t')\label{cond-2}
\end{align}

We use the Dirac Delta function integral representation to treat these order parameters as independent fields: as an example, for $\rho_n$ this amounts to inserting into $\mathcal{Z}$ the identity decomposition
\begin{align}
    1 = \prod_t\delta\bigg(\rho_n S - \sum_\sigma n_\sigma(t)\bigg) \propto \int~\mathcal{D}[\hat{\rho}] e^{iS \int~dt~\hat{\rho}_n(t)\rho_n(t)}e^{-i\sum_\sigma\int~dt~\hat{\rho}_n(t)~n_\sigma(t)}
\end{align}
Defining $\Pi = (\rho,\lambda,Q,L,K)$ and $\hat{\Pi} = (\hat{\rho},\hat{\lambda},\hat{Q},\hat{L},\hat{K})$, we finally write $Z$:
\begin{align}
    \mathcal{Z}[\psi,\phi] &= \int~\mathcal{D}[\Pi,\hat{\Pi}]~e^{S(\Psi[\Pi,\hat{\Pi}]+\Phi[\Pi])}\nonumber\\
    & \times\int~\mathcal{D}[n,\hat{n}]\mathcal{A}[n,\hat{n}]~e^{-i\sum_\sigma\int~dt[\hat{\rho}_n(t)n_\sigma(t)+\hat{\lambda}_n(t)\hat{n}_\sigma(t)]}e^{-i\sum_\sigma \int~dt\int~dt'~[\hat{Q}_n n_\sigma n_\sigma' +\hat{L}_n \hat{n}_\sigma \hat{n}_\sigma' +\hat{K}_n \hat{n}_\sigma n_\sigma']}\nonumber\\
    &\times \int~\mathcal{D}[c,\hat{c}]\mathcal{B}[c,\hat{c}]e^{-i\sum_j\int~dt[\hat{\rho}_c(t)c_j(t)+\hat{\lambda}_r(t)\hat{c}_j(t)]}e^{-i\sum_j\int~dt\int~dt'~[\hat{Q}_c c_j c_j' +\hat{L}_c \hat{c}_j \hat{c}_j' +\hat{K}_c \hat{c}_j c_j']}
\end{align}
where
\begin{align}\label{eq:z_before_saddle_point}
   \Psi[\Pi,\hat{\Pi}] &= i \int~dt~\big[\hat{\rho}_n(t)\rho_n(t)+\hat{\lambda}_n(t)\lambda_n(t)+\hat{\rho}_c(t)\rho_c(t)+\hat{\lambda}_r(t)\lambda_c(t)\big]\nonumber\\
   &+i\int~dt\int~dt'~[\hat{Q}_n ~Q_n+\hat{L}_n ~L_n+\hat{Q}_c ~Q_c+\hat{L}_c ~L_c+\hat{K}_n ~K_n+\hat{K}_c ~K_c](t,t')\\
   \Phi[\Pi]& = i \bar{\alpha}\int~dt~[\rho_{n}(t)\lambda_c(t)-\rho_c(t)\lambda_n(t)]\nonumber\\
  & -\frac{\Sigma^2}{2}\int~dt~\int~dt'\big[Q_n(t,t')L_c(t,t')+Q_c(t,t')L_n(t,t')-2K_n(t,t')K_c(t,t')\big] q(|t-t'|)
\end{align}

Now $\mathcal{Z}$ can be written in a compact form:
\begin{align}
   \mathcal{Z}[\psi,\phi] &= \int~\mathcal{D}[\Pi,\hat{\Pi}]~e^{S(\Psi[\Pi,\hat{\Pi}]+\Phi[\Pi]+\Omega_n[\hat{\Pi}]+\Omega_c[\hat{\Pi}])} \ .\label{z-compact}
\end{align}

To identify $\Omega_n[\hat\Pi]$, we start from the second line of Eq.~\eqref{eq:z_before_saddle_point}. We first write $e^{\sum_\sigma(\dots)} = \prod_{\sigma}e^{(\dots)}$, and then we factorize the path integral over the $\sigma$ indices, since these integrals are not coupled and can be carried out independently. Finally, we turn the product into a summation through the logarithm function. Since each term inside the summation is identical, the prefactor $1/S$ cancels out, we suppress indices $\sigma$ and define a unique auxiliary fields $h_n(t)$, obtaining (we omit some function arguments for ease of notation, using shorthands such as $n(t)n(t') \to n n'$)
\begin{align}
    \Omega_n[\hat{\Pi}]
    &=\ln\int~\mathcal{D}[n,\hat{n}] \exp\bigg[i \int~dt~\psi(t)n(t)+i \int~dt~\hat{n}(t)\bigg(\dfrac{\dot n(t)}{n(t)} + \delta-h_n(t)\bigg)\bigg]\nonumber\\
    &\times \exp\bigg[-i\int~dt[\hat{\rho}_n(t)n(t)+\hat{\lambda}_n(t)\hat{n}(t)]-i\int~dt\int~dt'~[\hat{Q}_n n n' +\hat{L}_n \hat{n} \hat{n}' +\hat{K}_n \hat{n} n']\bigg] \ .
\end{align}

In a similar manner we obtain the following object:
\begin{align}
  \Omega_c[\hat{\Pi}] &=\dfrac{1}{\nu}\ln\int~\mathcal{D}[c,\hat{c}] \exp\bigg[i \int~dt~\phi(t)c(t)+i \int~dt~\hat{c}(t)\bigg(\dfrac{\dot c(t)}{c(t)} - \mu[1- c(t)/\kappa]-h_c(t)\bigg)\bigg]\nonumber\\
    &\times \exp\bigg[-i\int~dt[\hat{\rho}_c(t)c(t)+\hat{\lambda}_r(t)\hat{c}(t)]-i\int~dt\int~dt'~[\hat{Q}_c c c' +\hat{L}_c \hat{c} \hat{c}' +\hat{K}_c \hat{c} c']\bigg]
\end{align}
Here, $\nu = S/R$ appears as sums over the $i$ indices give $R$ identical contributions.   

$\mathcal{Z}$, as given in Eqs.~\eqref{z-compact}, can be approximated using the saddle-point method for large $S$ and $R$, with fixed $\nu$:
\begin{align}
    \mathcal{Z}\approx e^{S(\Psi[\Pi^*,\hat{\Pi^*}]+\Phi[\Pi^*]+\Omega_n[\hat{\Pi^*}]+\Omega_c[\hat{\Pi^*}])}, \label{sad-z}
\end{align}
where the fields marked by $*$ solve the saddle point equations:
\begin{align}
    &\dfrac{\delta \Psi}{\delta \Pi}+\dfrac{\delta \Phi}{\delta \Pi} = 0 \ ,\label{ex-1}\\
    &\dfrac{\delta \Psi}{\delta \hat{\Pi}}+\dfrac{\delta \Omega_n}{\delta \hat{\Pi}}+\dfrac{\delta \Omega_c}{\delta \hat{\Pi}} = 0 \ .~\label{ex-2}
\end{align}
Eq.~\eqref{ex-1} is solved by
\begin{align}
    \hat{\rho}_n(t) = -\bar{\alpha} \lambda_c(t) = 0 &&\qquad\qquad  \hat{\lambda}_n(t) = \bar{\alpha}\rho_c(t)\label{0-eqns}\\
    \hat{\rho}_c(t) = \bar{\alpha} \lambda_n(t) = 0 && \hat{\lambda}_r(t) = -\bar{\alpha} \rho_n(t)\\
    i\hat{Q}_n =\frac{\Sigma^2}{2}L_c ~q(|t-t'|) = 0 && i\hat{L}_n =\frac{\Sigma^2}{2}Q_cq(|t-t'|)\\
    i\hat{Q}_c =\frac{\Sigma^2}{2}L_n  ~q(|t-t'|)  = 0   && i\hat{L}_c =\frac{\Sigma^2}{2}Q_nq(|t-t'|)\\
    i\hat{K}_n =-\Sigma^2 K_c ~q(|t-t'|)       && i\hat{K}_c =-\Sigma^2 K_nq(|t-t'|)
\end{align}

Eq.~\eqref{ex-2} instead gives
\begin{align}
    \rho_n(t) = \langle n \rangle_{\Omega_n} &&\qquad\qquad\lambda_n(t) =\langle \hat{n} \rangle_{\Omega_n} \label{pf-eq-1}\\
    \rho_c(t) = \dfrac{1}{\nu}\langle c \rangle_{\Omega_c}&&\lambda_c(t) =\dfrac{1}{\nu}\langle \hat{c}\rangle_{\Omega_c} \\
    Q_n = \langle nn'\rangle_{\Omega_n}&& L_n = \langle \hat{n}\hat{n}'\rangle_{\Omega_n}\\
    K_n = \langle \hat{n}n'\rangle_{\Omega_n}&& Q_c = \dfrac{1}{\nu}\langle cc'\rangle_{\Omega_c}\\
    L_c = \dfrac{1}{\nu}\langle \hat{c}\hat{c}'\rangle_{\Omega_c}&& K_c = \dfrac{1}{\nu}\langle \hat{c}c'\rangle_{\Omega_c}
\end{align}

In the equalities above, $\langle \cdots\rangle_{\Gamma}$ signifies averaging with respect to the \textit{effective action} $\Gamma = \{\Omega_n,\Omega_c\}$.
In Appendices~\ref{pf-A} and \ref{pf-B} we prove some of these equalities.

Using the above identities, which are valid in the saddle point approximation  (and exact in the limit $S,R\to\infty$), we first write $\Omega_n[\hat\Pi]$:
\begin{align}
    \Omega_n[\hat{\Pi}] &=\ln\int~\mathcal{D}[n,\hat{n}] \exp\bigg[i \int~dt~\psi(t)n(t)+i \int~dt~\hat{n}(t)\bigg(\dfrac{\dot n(t)}{n(t)} + \delta -h_n(t)\bigg)\bigg]\nonumber\\
    &\times \exp\bigg[-i\int~dt~\bar{\alpha}\rho_c(t)\hat{n}(t)-i\int~dt\int~dt'~[-i\frac{\Sigma^2}{2}Q_c \hat{n} \hat{n}'q(|t-t'|) +i~\hat{n}\Sigma K_c n' q(|t-t'|)]\bigg]\\
    &=\ln\int~\mathcal{D}[n,\hat{n}] \exp\bigg[i \int~dt~\psi(t)n(t)+i \int~dt~\hat{n}(t)\bigg(\dfrac{\dot n(t)}{n(t)} + \delta-h_n(t)-~\bar{\alpha}\rho_c(t)-i\Sigma^2\int~dt'~K_c(t,t')n(t')q(|t-t'|)\bigg)\nonumber\\&
    \times\exp\bigg[-\frac{\Sigma^2}{2}\int~dt\int~dt'~[\hat{n}(t)Q_c(t,t')\hat{n}(t')q(|t-t'|)]\bigg] \label{omega_n-eqn}
\end{align}
This gives the effective equation for $n(t)$:
\begin{align}
    \dfrac{\dot n(t)}{n(t)} =\frac{\bar{\alpha}\langle c \rangle(t)}{\nu} -\delta+i\Sigma^2\int~dt'~K_c(t,t')n(t')q(|t-t'|) +h_n(t)+\dfrac{\Sigma}{\sqrt{\nu}}\xi_n(t)
\end{align}
where the noise has the following correlation:
\begin{align}
    \langle \xi_n(t)\xi_n(t')\rangle = \sqrt{\nu}Q_c(t,t')q(|t-t'|) = q(|t-t'|)\langle c(t)c(t') \rangle 
\end{align}
Further, redefining the response function $G_c(t,t')/\nu=-iK_c(t,t')$, we get
\begin{align}
    \dfrac{\dot n(t)}{n(t)} =\frac{\bar{\alpha}\langle c \rangle(t)}{\nu}-\delta+h_n(t)+\dfrac{\Sigma}{\sqrt{\nu}}\xi_n(t)-\dfrac{\Sigma^2}{\nu}\int~dt'~q(|t-t'|)~G_c(t,t')n(t')
\end{align}
Similarly, we write 
\begin{align}
  \Omega_c[\hat{\Pi}] &=\dfrac{1}{\nu}\ln\int~\mathcal{D}[r,\hat{c}] \exp\bigg[i \int~dt~\phi(t)c(t)+i \int~dt~\hat{c}(t)\bigg(\dfrac{\dot c(t)}{c(t)} - \mu[1- c(t)/\kappa]-h_c(t)\bigg)\bigg]\nonumber\\
    &\times \exp\bigg[-i\int~dt[-\bar{\alpha}\rho_n\hat{c}(t)]-i\int~dt\int~dt'~[-i\frac{\Sigma^2}{2}Q_n\hat{c} \hat{c}' +i\Sigma^2 K_n\hat{c} r']q(|t-t'|)\bigg]\nonumber\\
    &=\dfrac{1}{\nu}\ln\int~\mathcal{D}[r,\hat{c}] \exp\bigg[i \int~dt~\phi(t)c(t)~+ \nonumber\\
    &+i \int~dt~\hat{c}(t)\bigg(\dfrac{\dot c(t)}{c(t)} - \mu[1- c(t)/\kappa]-h_c(t)+\bar{\alpha}\rho_n-i\Sigma^2\int~dt'~K_n(t,t')c(t')q(|t-t'|)\bigg)\bigg]\nonumber\\
    &\times \exp\bigg[-i\int~dt\int~dt'~[-i\frac{\Sigma^2}{2}Q_n\hat{c} \hat{c}'q(|t-t'|)]\bigg]\label{omega_r-eqn}
\end{align}
Thus, the effective equation for $c(t)$ is 
\begin{align}
    \dfrac{\dot c(t)}{c(t)} = \mu[1- c(t)/\kappa]- \bar{\alpha}\langle n \rangle (t) +h_c(t)+ \Sigma\xi_c(t) +i\Sigma^2\int~dt'~q(|t-t'|)K_n(t,t')c(t') 
\end{align}
with noise correlations:
\begin{align}
    \langle \xi_c(t)\xi_c(t')\rangle = Q_n(t,t') q(|t-t'|) = q(|t-t'|) \langle n(t)n(t') \rangle 
\end{align}
We redefine the response function $G_n(t,t')=-iK_n(t,t')$ (Appendix~\ref{sec:response}), and we get
\begin{align}
    \dfrac{\dot c(t)}{c(t)} = \mu[1- c(t)/\kappa]- \bar{\alpha}\langle n \rangle (t) +h_c(t)+ \Sigma\xi_c(t) -\Sigma^2\int~dt'~q(|t-t'|)G_n(t,t')c(t') 
\end{align}

Let us write the effective DMFT equations, which coincide with Eqs.~\eqref{eq:DMFT_effective_dynamics} of the main text, provided we perform the substitution $\Sigma^2 q(|t-t'|) \to \Delta^2 + \Sigma^2q(|t-t'|)$ as anticipated:
\begin{align}
        \dfrac{\dot n(t)}{n(t)} &=\bar{\alpha}\langle c \rangle(t)-\delta+\dfrac{\Sigma}{\sqrt{\nu}}\xi_n(t)-\dfrac{\Sigma^2}{\nu}\int_0^t~dt'~q(|t-t'|)~G_c(t,t')n(t') + h_n(t) \label{eff-1}\\
     \dfrac{\dot c(t)}{c(t)}& = \mu[1- c(t)/\kappa]- \bar{\alpha}\langle n \rangle(t) + \Sigma\xi_c(t) -\Sigma^2\int_0^t~dt'~q(|t-t'|)G_n(t,t')c(t')  + h_c(t)\label{eff-2}
\end{align}

\begin{align}
        \langle \xi_n(t)\xi_n(t')\rangle = q(|t-t'|)\langle c(t)c(t') \rangle \\
    \langle \xi_c(t)\xi_c(t')\rangle = q(|t-t'|) \langle n(t)n(t') \rangle 
\end{align}
where $q(|t-t'|)$ is given in Eq.~\eqref{z-corr-t}, and the response functions are calculated as (see Appendix~\ref{sec:response})

\begin{align}
G_{n}(t,t') = \bigg\langle \dfrac{\delta n(t)}{\delta h_n(t')} \bigg\rangle \ , \quad G_{c}(t,t') = \bigg\langle \dfrac{\delta c(t)}{\delta h_c(t')} \bigg\rangle \ .
\end{align}

\subsection{Proof of Eq.~\eqref{pf-lastline}}\label{sec:lastline}
Here we give the proof of Eq.~\eqref{pf-lastline}:
\begin{align}
    \overline{\exp\bigg[\frac{i\Sigma}{\sqrt{S}} \int_0^t dt~A(t)Z(t)\bigg]} &= 1 - \frac{\Sigma^2}{2!S} \int_0^t dt_1\int_0^t dt_2~A(t_1)A(t_2)\overline{Z(t_1)Z(t_2)}+\nonumber\\
    &+  \frac{\Sigma^4}{4!S^2} \int_0^t dt_1\int_0^t dt_2\int_0^t dt_3\int_0^t dt_4 ~A(t_1)A(t_2)A(t_3)A(t_4)\overline{Z(t_1)Z(t_2)Z(t_3)Z(t_4)} + \dots\ , \label{exp-fun}
\end{align}
8where we have dropped terms which are odd in $Z$, since these will be zero after averaging over the Gaussian distribution $P(Z)$. $Z$ is a stationary colored noise with two-time correlation: $\overline{Z(t_1)Z(t_2)} = q(|t_1-t_2|)$.

To evaluate the r.h.s. of Eq.~\eqref{exp-fun}, we make use of Wick's theorem for Gaussian random variables:
\begin{align}
   \overline{Z(t_1)Z(t_2)} &= q(|t_1-t_2|) \\
   \overline{Z(t_1)Z(t_2)Z(t_3)Z(t_4)} &= \overline{Z(t_1)Z(t_2)}~~\overline{Z(t_3)Z(t_4)} + \overline{Z(t_1)Z(t_3)}~~\overline{Z(t_2)Z(t_4)}+\overline{Z(t_1)Z(t_4)}~~\overline{Z(t_2)Z(t_3)}\ ,\label{wick}
\end{align}

through which we get
\begin{align}
    \overline{\exp\bigg[\frac{i\Sigma}{\sqrt{S}} \int_0^t dt~A(t)Z(t)\bigg]} &= 1 - \frac{\Sigma^2}{2!S} \int_0^t dt_1\int_0^t dt_2~A(t_1)A(t_2)q(|t_1-t_2|)+\nonumber\\
    &+  \frac{3\Sigma^4}{4!S^2} \int_0^t dt_1\int_0^t dt_2\int_0^t dt_3\int_0^t dt_4 ~A(t_1)A(t_2)A(t_3)A(t_4)q(|t_1-t_2|)q(|t_3-t_4|) + \dots\ , \nonumber\\
    &= 1 - \frac{\Sigma^2}{2S} \int_0^t dt_1\int_0^t dt_2~A(t_1)A(t_2)q(|t_1-t_2|)+\nonumber\\
    &+  \dfrac{1}{2!} \bigg(\frac{\Sigma^2}{2S}\bigg)^2\bigg[\int_0^t dt_1\int_0^t dt_2 ~A(t_1)A(t_2)q(|t_1-t_2|)\bigg]^2 + \dots\\
    & = \exp\bigg[-\dfrac{\Sigma^2}{2S} \int_0^t dt_1\int_0^t dt_2 ~A(t_1)A(t_2)q(|t_1-t_2|)\bigg]\ .
\end{align}

The above result is used to evaluate the last line of Eq.~\eqref{last-line}.

\subsection{Proof of Eq.~\eqref{pf-eq-1}}\label{pf-A}
The left part of Eq.~\eqref{pf-eq-1} reads:
\begin{align}
    \rho_n(t) = \langle n \rangle_{\Omega_n}.\label{app-1}
\end{align}
This equation is obtained from Eq~\eqref{ex-2}. In the following, we give the detail of the derivation of Eq.~\eqref{app-1}. 

From Eq.~\eqref{ex-2}, we have
\begin{align}
    \dfrac{\delta\Psi}{\delta\hat{\Pi}}+\dfrac{\delta\Omega_n}{\delta\hat{\Pi}}+\dfrac{\delta\Omega_c}{\delta\hat{\Pi}}=0
\end{align}
Let us first consider $\hat{\Pi}(t') = \hat{\rho}_n(t')$, then the first term gives
\begin{align}
    \dfrac{\delta\Psi}{\delta \hat{\rho}_n(t')}=i\int~dt~\delta(t-t')\rho_n(t) = i \rho_n(t')\label{sum-1}
\end{align}
Similarly, the second term gives
\begin{align}
    \dfrac{\delta\Omega_n}{\delta \hat{\rho}_n(t')}&= \dfrac{\int~\mathcal{D}[n,\hat{n}] \cdots e^{-i\int~dt[\hat{\rho}_n(t)n(t)+\hat{\lambda}_n(t)\hat{n}(t)]} (-i\frac{\partial}{\partial \hat{\rho}(t')})\int~dt~\hat{\rho}_n(t)n(t) \cdots}{\int~\mathcal{D}[n,\hat{n}] e^{i \int~dt~\psi(t)n(t)+i \int~dt~\hat{n}(t)\bigg(\frac{\dot n(t)}{n(t)} + \delta -h_n(t)\bigg)}~e^{-i\int~dt[\hat{\rho}_n(t)n(t)+\hat{\lambda}_n(t)\hat{n}(t)]-i\int~dt\int~dt'~[\hat{Q}_n n n' +\hat{L}_n \hat{n} \hat{n}' +\hat{K}_n \hat{n} n']}}\nonumber\\
   & = \dfrac{\int~\mathcal{D}[n,\hat{n}] \cdots e^{-i\int~dt[\hat{\rho}_n(t)n(t)+\hat{\lambda}_n(t)\hat{n}(t)]} (-i)\int~dt~\delta(t-t')n(t) \cdots}{\int~\mathcal{D}[n,\hat{n}] e^{i \int~dt~\psi(t)n(t)+i \int~dt~\hat{n}(t)\bigg(\frac{\dot n(t)}{n(t)} + \delta -h_n(t)\bigg)}~e^{-i\int~dt[\hat{\rho}_n(t)n(t)+\hat{\lambda}_n(t)\hat{n}(t)]-i\int~dt\int~dt'~[\hat{Q}_n n n' +\hat{L}_n \hat{n} \hat{n}' +\hat{K}_n \hat{n} n']}}\nonumber\\
    &=-i\langle n(t') \rangle_{\Omega_n}\label{sum-2}
\end{align}
The third term gives zero. Summing the two contributions in Eqs.~\eqref{sum-1} and \eqref{sum-2} we obtain \eqref{app-1}.
We can compute the expressions of the other order parameters in a similar fashion.

\subsection{Proof of Eq.~\eqref{0-eqns}}\label{pf-B}
In this section, we give a proof of the leftmost part of Eq.~\eqref{0-eqns}:
\begin{align}
    \hat{\rho}_n(t) = -\bar{\alpha} \lambda_c(t) = 0,
\end{align}
where $\lambda_c=0$. 

Let us look at the equation for $\lambda_c(t):$
\begin{align}
    \lambda_c(t) = \dfrac{1}{S}\sum_j~\hat{c}_j(t).
\end{align}
We will show that $\hat{c}_j=0$ for all $j$.
\begin{align}
    Z[\psi,\phi] &= \int~\mathcal{D}[n,\hat{n}]~\mathcal{D}[r,\hat{c}]~\mathcal{A}[n,\hat{n}]\Delta[n,\hat{n},r,\hat{c}]\nonumber\\
    &\times\exp\bigg[i\sum_j \int~dt~\phi_j(t)c_j(t)+i\sum_j \int~dt~\hat{c}_j(t)\bigg(\dfrac{\dot{c}_j(t)}{c_j(t)}+ \mu[1-c_j(t)/\kappa]-h_j(t)\bigg)\bigg].
\end{align}
Setting $\phi\to 0$ and $\phi\to 0$ gives $Z[\psi=0,\phi=0]=1$ (by definition).
Differentiating $Z[\psi=0,\phi=0]$ with respect to $h_k(t')$ gives:
\begin{align}
    \underbrace{\dfrac{\delta Z[\psi=0,\phi=0]}{\delta h_k(t')}}_{=0}=\int~\mathcal{D}[n,\hat{n}]~\mathcal{D}[r,\hat{c}]~\mathcal{A}[n,\hat{n}]\Delta[n,\hat{n},r,\hat{c}]~e^{i\sum_j \int~dt~\hat{c}_j(t)\bigg(\dfrac{\dot{c}_j(t)}{c_j(t)}+ a^\mu~c_j(t)-h_j(t)\bigg)} [ -i \hat{c}_k(t')],
\end{align}
where the term in the square bracket is obtained as we showed in the previous section~\eqref{pf-A}.
Since the left-hand side is identically equal to 0 for any choice of $h_k(t')$ in the functional derivative, and that without the factor $ -i \hat{c}_k(t')$ the integral would evaluate to 1, it has to hold $\hat{c}_k(t')=0$. By same token, if we differentiate with respect to $\hat{c}_{k}(t')$, we get the equation of motion for resources in the presence of field $h_{k}(t)$. Therefore, $\lambda_c(t)=0$. The fact that other order parameters are identically equal to 0 can be shown with analogous calculations. 

\subsection{Response function}\label{sec:response}
Here we compute the response functions of the effective dynamics. We begin by considering
\begin{align}
    \Omega_n[\hat{\Pi}] &=\ln Z_{n,1}[\psi]\ , \label{omega_n-eqn2}
\end{align}
where
\begin{align}
    Z_{n,1}[\psi]&\equiv \int~\mathcal{D}[n,\hat{n}] \exp\bigg[i \int~dt~\psi(t)n(t)+i \int~dt~\hat{n}(t)\bigg(\dfrac{\dot n(t)}{n(t)} + \delta -h_n(t)-~\bar{\alpha}\rho_c(t)-i\Sigma^2\int~dt'~K_c(t,t')n(t')q(|t-t'|)\bigg)\nonumber\\&
    \times\exp\bigg[-\frac{\Sigma^2}{2}\int~dt\int~dt'~[\hat{n}(t)Q_c(t,t')\hat{n}(t')q(|t-t'|)]\bigg]\ .
\end{align}
Notice that $Z[\psi = 0]=1$ for any given choice of $h_n(t)$. 

Differentiating Eq.~\eqref{omega_n-eqn2} with respect to $\psi(t)$, we get
\begin{align}
     \dfrac{\delta \Omega_n[\hat{\Pi}]}{\delta \psi(t)} &=\dfrac{1}{Z_{n,1}[\psi]} \dfrac{\delta Z_{n,1}[\psi]}{\delta \psi(t)}
\end{align}

To compute $\langle n(t) \rangle$, we set $\psi = 0$, and this gives
\begin{align}
\dfrac{\delta \Omega_n[\hat{\Pi}]}{\delta \psi(t)}\bigg|_{\psi = 0}= i \langle n(t) \rangle\ , \label{mom-n}
\end{align}
where 
\begin{align}
    \langle n(t) \rangle &= \int~\mathcal{D}[n,\hat{n}] ~n(t)~\exp\bigg[i \int~dt~\hat{n}(t)\bigg(\dfrac{\dot n(t)}{n(t)} + \delta -h_n(t)-~\bar{\alpha}\rho_c(t)-i\Sigma^2\int~dt'~K_c(t,t')n(t')q(|t-t'|)\bigg)\nonumber\\&
    \times\exp\bigg[-\frac{\Sigma^2}{2}\int~dt\int~dt'~[\hat{n}(t)Q_c(t,t')\hat{n}(t')q(|t-t'|)]\bigg]
\end{align}
Here the average is performed over the effective dynamics~\eqref{eff-1}.

Now, differentiating with respect to $h_n(t')$, we get
\begin{align}
    G_n(t,t')\equiv \dfrac{\delta \langle n(t) \rangle }{\delta h_n(t')} = -i \langle n(t)\hat{n}(t') \rangle \equiv - i K_n(t,t')\ .
\end{align}

Similarly, consider the response function for the resource:
\begin{align}
    \Omega_c[\hat{\Pi}] &=\dfrac{1}{\nu}\ln Z_{r,1}[\phi] \label{omega_r-eqn2}
\end{align}
where
\begin{align}
   Z_{r,1}[\phi]&=\int~\mathcal{D}[r,\hat{c}] \exp\bigg[i \int~dt~\phi(t)c(t)~+ \nonumber\\
    &+i \int~dt~\hat{c}(t)\bigg(\dfrac{\dot c(t)}{c(t)} - \mu[1- c(t)/\kappa]-h_c(t)+\bar{\alpha}\rho_n-i\Sigma^2\int~dt'~K_n(t,t')c(t')q(|t-t'|)\bigg)\bigg]\nonumber\\
    &\times \exp\bigg[-i\int~dt\int~dt'~[-i\frac{\Sigma^2}{2}Q_n\hat{c} \hat{c}'q(|t-t'|)]\bigg]
\end{align}
As before, $Z_{n,1}[\phi = 0]=1$ for a given auxiliary field $h_c(t)$. 

Differentiating Eq.~\eqref{omega_r-eqn2} with respect to $\phi(t)$, we get
\begin{align}
     \dfrac{\delta \Omega_n[\hat{\Pi}]}{\delta \phi(t)} &=\dfrac{1}{\nu}\dfrac{1}{Z_{n,1}[\psi]} \dfrac{\delta Z_{n,1}[\psi]}{\delta \psi(t)}
\end{align}

To compute $\langle c(t) \rangle$, we set $\phi = 0$:
\begin{align}
\dfrac{\delta \Omega_c[\hat{\Pi}]}{\delta \phi(t)}\bigg|_{\phi = 0}= \dfrac{1}{\nu}i \langle c(t) \rangle, \label{mom-n}
\end{align}
where
\begin{align}
   \langle c(t) \rangle&=\int~\mathcal{D}[r,\hat{c}] c(t)\exp\bigg[i \int~dt~\hat{c}(t)\bigg(\dfrac{\dot c(t)}{c(t)} - \mu[1- c(t)/\kappa]-h_c(t)+\bar{\alpha}\rho_n-i\Sigma^2\int~dt'~K_n(t,t')c(t')q(|t-t'|)\bigg)\bigg]\nonumber\\
    &\times \exp\bigg[-\dfrac{\Sigma^2}{2}\int~dt\int~dt'~Q_n\hat{c} \hat{c}'q(|t-t'|)\bigg]\ ,\label{omega_r-eqn3}
\end{align}
Here the average is performed over the effective dynamics~\eqref{eff-2}.

Differentiating Eq.~\eqref{omega_r-eqn3} with respect to $h_c(t')$, we get
\begin{align}
    G_c(t,t')\equiv \dfrac{\delta \langle c(t) \rangle }{\delta h_c(t')} = -i \langle c(t)\hat{c}(t') \rangle \equiv - i \nu K_c(t,t')\ .
\end{align}

\section{Stationary distributions of the effective dynamics in the white noise limit}
\label{si:WNL}

The effective dynamics~\eqref{eq:DMFT_effective_dynamics} are valid for any choice of $\tau$, which enters the definition of the temporal correlation of the noise, namely $q(|t-t'|) = \frac{1+2\tau}{2\tau}e^{-\frac{|t-t'|}{\tau}}$. In the steady state, i.e., for $t \to +\infty$, all two-time quantities (response functions and correlations) depend
solely on the time difference $t - t'$.
Furthermore, in the limit of vanishing correlation time, $\tau \to 0$, the two noises $\xi_{n,c}(t)$ 
appearing in~\eqref{eq:DMFT_effective_dynamics} have correlation structures which are local in time:
$\langle \xi_n(t)\xi_n(t')\rangle = \big\langle \big[c(t)\big]^2\big\rangle\delta(t-t')$ and $\langle \xi_c(t)\xi_c(t')\rangle = \big\langle \big[n(t)\big]^2\big\rangle\delta(t-t')$. Using this, we calculate the response functions:
\begin{equation}\label{eq:WNL_cesponse_function}
G_n(t,t') \equiv \bigg\langle \frac{\delta n(t)}{\delta h_n(t')} \bigg\rangle =
\bigg\langle \frac{\delta}{\delta h_n(t')} \cdots \int_0^t dt' n(t')h_n(t')  \cdots  \bigg\rangle = \langle n(t') \rangle \Theta(t-t')\ ,
\end{equation}
so that, as $t' \to t$, 
\begin{align}
    G_n(t,t') \to \frac{1}{2}\langle n(t) \rangle\ ,\label{eq:gn}
\end{align}
where the factor $\frac{1}{2}$ on the right-hand side is due to the definition of the step function $\Theta(\cdot)$, which has to be consistent with the {\it Stratonovich interpretation} of the stochastic equations~\eqref{eq:DMFT_effective_dynamics}. Similarly, one finds the response function for the resource concentration as
\begin{align}
    G_c(t,t') \to \frac{1}{2}\langle c(t) \rangle\ .\label{eq:gc}
\end{align} 
Taking into account the above Eqs.~\eqref{eq:gn} and \eqref{eq:gc}, we set the auxiliary fields $h_n = h_c = 0$, and rescale the noise terms to make explicit their dependence on $\langle n \rangle$ and $\langle c \rangle$. Further, changing from Stratonovich calculus to Ito calculus by introducing suitable terms \cite{gardiner2004handbook}, we rewrite \eqref{eq:DMFT_effective_dynamics} as
\begin{align} 
\label{eq:WNL_effective_dynamics}
\dot{n} &= n\Big(\frac{\bar\alpha}{\nu}\langle c \rangle - \delta + \frac{\Sigma\sqrt{\langle c^2 \rangle}}{\sqrt{\nu}}\xi_n - \frac{\Sigma^2}{2\nu}\langle c \rangle n \Big) + \lambda_n \ , \\
\dot{c} &= \mu c \Big(1 - \frac{c}{\kappa}\Big) - c\Big(\bar\alpha\langle n \rangle + \Sigma\sqrt{\langle n^2 \rangle} \xi_c + \frac{\Sigma^2}{2}\langle n \rangle c \Big) + \lambda_c\ ,
\end{align}
where the noises $\xi_x(t)$ have zero mean, and are delta correlated in time: $\langle \xi_x(t) \xi_y(t') \rangle = \delta_{x,y}\delta(t-t')$, for $x,y \in \{n,c\}$. 
From these, it is straightforward to write down the stationary distributions of both $n$ and $c$:
\begin{align}\label{eq:ss_SAD_SI}
p^{(n)}_{\rm st}(n) &= Z_n^{-1} n^{-1+d_n} e^{-n/b_n}e^{-\ell_n/n}\ , \\
p^{(c)}_{\rm st}(c) &= Z_c^{-1} c^{-1+d_c} e^{-c/b_c}e^{-\ell_c/c}\ ,
\end{align}
where we defined $d_n \equiv 2\dfrac{\bar\alpha\langle c \rangle - \nu \delta}{\Sigma^2 \langle c^2 \rangle}$, $b_n \equiv \dfrac{\langle c^2 \rangle}{\langle c \rangle}$, $\ell_n \equiv 2\dfrac{\nu\lambda_n}{\Sigma^2\langle c^2 \rangle}$, $d_c \equiv 2 \dfrac{\mu - \bar \alpha \langle n \rangle}{\Sigma^2 \langle n ^2 \rangle}$, 
$b_c \equiv \dfrac{\Sigma^2\langle n ^ 2 \rangle}{2\mu/\kappa + \Sigma^2 \langle n \rangle}$, $\ell_c \equiv 2\frac{\lambda_c}{\Sigma^2\langle n^2 \rangle}$; $Z_n$ and $Z_c$ are normalization constants.
Let us set $\lambda_n = 0 = \lambda_c$, so that the distributions in Eqs.~\eqref{eq:ss_SAD_SI} become Gamma distributions. In this case we can write closed equations for the $\langle n \rangle$ and $\langle c \rangle$, namely
\begin{align}\label{eq:ss_SAD_1st_moments}
    \langle n \rangle &= d_n b_n = 2 \frac{\bar\alpha\langle c \rangle - \nu \delta}{\Sigma^2 \langle c \rangle} \ , \\
    \langle c \rangle &= d_c b_c = 2\frac{\mu - \bar\alpha\langle n \rangle}{2\mu/\kappa + \Sigma^2 \langle n \rangle}  \ ,    
\end{align}
to which a unique positive solution exists, provided that $\nu > \nu_c \equiv \bar\alpha \kappa / \delta$:
\begin{align}\label{eq:ss_SAD_selfcon_means}
    \langle n \rangle &= \frac{2 \bar{\alpha}^2 \kappa +\kappa  \Sigma ^2 (\delta  \nu +\mu )-\sqrt{\kappa  \left(4 \bar{\alpha}^4 \kappa +4 \bar{\alpha}  \Sigma ^2 (\bar{\alpha}  \delta  \kappa  \nu -\bar{\alpha}  \kappa  \mu +2 \delta  \mu  \nu )+\kappa  \Sigma ^4 (\delta  \nu +\mu )^2\right)}}{2 \bar{\alpha}  \kappa  \Sigma ^2} \ , \\
    \langle c \rangle &= \frac{-2 \bar{\alpha}^2 \kappa +\kappa  \Sigma ^2 (\delta  \nu +\mu )+\sqrt{\kappa  \left(4 \bar{\alpha}^4 \kappa +4 \bar{\alpha}  \Sigma ^2 (\bar{\alpha}  \delta  \kappa  \nu -\bar{\alpha}  \kappa  \mu +2 \delta  \mu  \nu )+\kappa  \Sigma ^4 (\delta  \nu +\mu )^2\right)}}{2 \Sigma ^2 (\bar{\alpha}  \kappa +\mu )} \ .
\end{align}
Having found the first moments, we obtain closed equations for the second moments:
\begin{align}\label{eq:ss_SAD_2nd_moments}
    \langle n^2 \rangle - \langle n \rangle^2 &= d_n b_n^2 = \frac{\langle n \rangle}{\langle c \rangle}\langle c^2 \rangle \ , \\
    \langle c^2 \rangle - \langle c \rangle^2 &= d_c b_c^2 = \frac{\langle c \rangle}{\frac{2\mu}{\kappa \Sigma^2}+\langle n \rangle}\langle n^2 \rangle \ ,
\end{align}
which can be uniquely solved for $\langle n ^2 \rangle$ and $\langle c^2 \rangle$, both found to be positive. Thus, we are able to write down the parameters of the stationary distributions of $n$ and $c$ in terms of the model's parameters; in particular, the parameters of $p_{\rm st}^{(n)}$ have relatively simple expressions, reading

\begin{align}\label{eq:wnl_gamma_parameters}
d_n = \frac{A}{2\mu \Sigma^2 \nu_c}\frac{1}{b_n} ,  \quad b_n = \frac{\delta \nu}{\bar\alpha}\Big(1 + \frac{\delta \nu + \mu}{4\bar\alpha \delta \nu \mu}A\Big)\ ,
\end{align}
for 
\begin{equation}
A \equiv \kappa(2 \bar\alpha ^2+\Sigma ^2 (\delta  \nu +\mu )) -\sqrt{\kappa (4 \bar\alpha ^4 \kappa+4 \bar\alpha  \Sigma ^2 (2 \delta  \mu  \nu +\bar\alpha  \kappa (\delta  \nu -\mu ))+\kappa \Sigma ^4 (\delta  \nu +\mu )^2)}\ .
\end{equation}

$A$ is found to be positive if $\nu < \nu_c \equiv \bar\alpha \kappa / \delta$. Having an explicit expression for $d_n$ is particularly useful when computing evenness, as shown in Appendix~\ref{si:evenness}.

\section{Stationary distributions of the effective dynamics for general correlation time}
\label{si:UCNA}
\renewcommand{\thefigure}{C\arabic{figure}}
\renewcommand{\thetable}{C\arabic{table}}
\setcounter{figure}{0}
\setcounter{table}{0}
Here we show the necessary steps to derive the approximated stationary distribution of the effective process, Eq.~\eqref{eq:DMFT_effective_dynamics}, for generic $\tau$. This procedure is possible because, as noticed in the main text, the effective dynamics of $n(t)$ and $c(t)$ in Eq.~\eqref{eq:DMFT_effective_dynamics} are only coupled in statistics, which become constants at stationarity; i.e.,   except for mutual self-consistency, the two equations are independent. We focus on $n(t)$; the corresponding result for $c(t)$ follows identical steps. 

Firstly, the noise which appear in the equation for $n(t)$ depends on the two-times correlation of $c(t)$. We check numerically that the correlation function of $c(t)$---and thus, the correlation of $\xi_n$--has (approximately, 
for short times) an exponential form. Since the stationary autocorrelation function should interpolate between $\langle c^2 \rangle$ at small times and $\langle c \rangle^2$ at large times, and calling $\tau_c$ the autocorrelation time of $c(t)$, we assume, for $|t-t'| \ll \tau_c$,

\begin{equation}\label{eq:autocorr_ansatz}
\langle c(t) c(t') \rangle 
\simeq
\big( \langle c ^ 2 \rangle - \langle c \rangle ^2 \big)e^{-|t-t'|/\tau_c} + \langle c \rangle ^2 \approx \langle c^2 \rangle e^{-|t-t'|/\tau_c'}\ ,
\end{equation}
where $\tau_c' = \frac{\langle c^2 \rangle}{\langle c^2 \rangle - \langle c \rangle ^2}\tau_c$. This enters the correlation function of $\xi_n(t)$, since

\begin{equation}\label{eq:noise_ansatz}
\langle \xi_n(t) \xi_n(t') \rangle / \Sigma^2 = q(t,t')\langle c(t) c(t') \rangle = \frac{2\tau}{1+2\tau}e^{-|t-t'|/\tau}\langle c^2 \rangle e^{-|t-t'|/\tau_c'} = \langle c^2 \rangle\frac{2\tau}{1+2\tau}e^{-|t-t'|/\bar\tau_n},
\end{equation}

with $\bar\tau_n = \big(\tau^{-1} + \tau_c'^{-1}\big)^{-1}$.\\

Secondly, we need to approximate the memory terms contained in \eqref{eq:DMFT_effective_dynamics} to a manageable form. We do this by assuming 
\begin{equation}\label{eq:memory_term_approximation}
\int_0^t G_c(t,t')q(t,t')n(t')dt' \equiv \chi_c n(t) \,
\end{equation}
where $\chi_c$ is a constant.
This assumption is exact in the limits $\tau \to 0$ and $\tau \to+\infty$, and in these cases the value of $\chi_c$ is known exactly -- explicitly for $\tau \to 0$ (see Appendix~\ref{si:WNL}), and as the numerical solution of a closed self-consistent equation for $\tau \to +\infty$ case \cite{BatistaToms2021}. However, in general, $\chi_c$ is left as a free parameter, to be determined by fitting the theoretical result to empirical distributions of abundances.

With the above assumptions, we can apply the results of \cite{Jung1987} in the case of a stochastic equation with colored multiplicative noise. We are considering a (Stratonovich) SDE of the form $\dot{n} = n(A + B n) + C n \epsilon$, where $\epsilon$ is an Ornstein-Uhlenbeck process, $\dot \epsilon = - \epsilon / \bar\tau + (\sqrt{2D}/\bar\tau_n) \xi$, and $\xi$ is Gaussian white noise, $\langle \xi \rangle = 0$ and $\langle \xi(t) \xi(t') \rangle = \delta(t-t')$. The constants just introduced are $A=\bar\alpha \langle c \rangle / \nu - \delta$, $B = \Sigma^2 \chi_c/\nu$, $C = \Sigma / \sqrt{\nu}$ and $D = \bar\tau_n \frac{1+2\tau}{2\tau}\langle c^2 \rangle$. Applying the result in note 14 of \cite{Jung1987} we obtain the first distribution (i.e., the SAD) of Eqs.~\eqref{eq:stat_distr_UCNA}.
In the same way, we obtain the second distribution of Eqs.~\eqref{eq:stat_distr_UCNA}, where $\tau_n$ is the autocorrelation time of $n(t)$, $\tau_n' = \frac{\langle n^2 \rangle}{\langle n^2 \rangle - \langle n \rangle^2}\tau_n$, $\bar\tau_c = (\tau^{-1} + \tau_n'^{-1})^{-1}$ and $\int_0^{t}G_n(t,t')q(t,t')c(t') \equiv \chi_n c(t)$.  All constants appearing in Eqs.~\eqref{eq:stat_distr_UCNA} are reported in Tab.~\ref{tab:ucna_constants}; normalization constants for both distributions can be computed analytically. It is straightforward to verify that the expressions above reduce to the white noise limit results, provided that as $\tau \to 0$ $\chi_c \to -\langle c \rangle / 2 $ and $\chi_n \to \langle n \rangle /2$. From this, we assume that $\chi_c <0$ and $\chi_n >0$ $\forall \tau$, which can be confirmed \textit{a posteriori} by means of numerical analysis.

\begin{table}[h!]
    \centering
    \begin{tabular}{c|cccc}
        \toprule
         & $A$ & $B$ & $C$ & $D$\\ 
        \midrule
        $n$ & $-\dfrac{\nu}{\bar{\tau}_n \Sigma^2 \chi_c}$ 
    & $\dfrac{\bar{\alpha} \langle c \rangle - \nu \delta}{\bar{\tau}_n \langle c^2 \rangle \Sigma^2}$ 
    & $\dfrac{1}{\langle c^2 \rangle \nu}$ 
    & $\chi_c \big[\nu (\delta + \bar{\tau}_n^{-1}) - \bar{\alpha} \langle c \rangle\big]$ \\ 
        & & & &\\
        $c$ & $\dfrac{1}{\bar{\tau}_c \left(\frac{\mu}{\kappa} + \Sigma^2 \chi_n \right)}$ 
    & $\dfrac{\mu - \bar{\alpha} \langle n \rangle}{\langle n \rangle^2 \Sigma^2 \bar{\tau}_c}$ 
    & $\dfrac{\mu/\kappa + \Sigma^2 \chi_n}{\langle n^2 \rangle \Sigma^2}$ 
    & $\dfrac{1}{\bar{\tau}_c} - \mu - \bar{\alpha} \langle n \rangle$ \\ 
        \bottomrule
    \end{tabular}
        \caption{Definitions of constants in Eqs.~\ref{eq:stat_distr_UCNA}. We highlight the fact that $\tau$ enters several of these constants, which also depend on the first and second moments of $n$ and $c$.}\label{tab:ucna_constants}
\end{table}

\section{Connection between empirical evenness and SAD}\label{si:evenness}

\renewcommand{\thefigure}{D\arabic{figure}}
\setcounter{figure}{0}

We consider an ecological community with species richness $S$ and (absolute) species abundances
$n_1,n_2,\dots, n_S$. The empirical evenness of this community is
\begin{align}
\dfrac{e^{H}}{S}\label{eq:even},
\end{align}
where $H$ stands for the Shannon entropy~\cite{Spellerberg2003} of the relative abundances $\{p_\sigma\}_\sigma$,
\begin{align}
    H = -\sum_{\sigma=1}^Sp_\sigma \ln p_\sigma .
\end{align}
We wish to derive an estimate of the evenness from the species abundance distribution (SAD) of the community, $P(n)$, in the limit of large $S$. To this end, we build a (categorical) distribution of relative abundances by sampling $S$ abundances from the SAD,
\begin{equation}
p_{\sigma} \equiv \frac{n_\sigma}{S\langle n \rangle}\ , \qquad \sigma = 1,\dots,S\ ,
\label{eq:relative_abundances_SAD}
\end{equation}
where $\langle n \rangle \equiv \int_0^{+\infty} dn~n P(n)$. Notice that, since we are treating the relative abundances as independent quantities, the constraint $\sum_\sigma p_\sigma = 1$ %will only be 
is only satisfied in the limit $S \to \infty$, given that $\frac{1}{S}\sum_\sigma n_\sigma \to \langle n \rangle$.  Within this construction, $e^H/S$ is a random variable which depends on the sample drawn from the SAD, therefore we compute the evenness by averaging over such samples: 
\begin{align}\label{eq:evenness_microscopic}
D &= \dfrac{1}{S} \langle e^{-\sum_{\sigma=1}^S p_\sigma \ln p_\sigma}\rangle \\
 &= \dfrac{1}{S} \bigg\langle \prod_{\sigma=1}^S e^{-\frac{n_\sigma}{S\langle n\rangle}\ln \frac{n_\sigma}{S\langle n\rangle}}\bigg\rangle\ .
\end{align}
Since each of $n_\sigma$ is independently and identically distributed according to the stationary SAD $P(n)$, we write 
\begin{align}
D &= \dfrac{1}{S} \prod_{\sigma=1}^S \bigg\langle e^{-\frac{n_\sigma}{S\langle n\rangle}\ln \frac{n_\sigma}{S\langle n\rangle}}\bigg\rangle\\
&= \dfrac{1}{S} \bigg\langle e^{-\frac{n}{S\langle n\rangle}\ln \frac{n}{S\langle n\rangle}}\bigg\rangle^S\\
&= \dfrac{1}{S} \bigg[\int_0^{+\infty}~dn~P(n)~e^{-\frac{n}{S\langle n\rangle}\ln \frac{n}{S\langle n\rangle}}\bigg]^S\ .
\end{align} 
Now, assume that the population is distributed according to a gamma distribution: $n \sim \text{Gamma}(d,b)$, so that $\langle n \rangle = d b$ and $\langle\langle n^2 \rangle\rangle = d b^2$. Then,
\begin{align}
    D&= \dfrac{1}{S} \bigg[\int_0^{+\infty}~dn~P(n)~\bigg(1 - \frac{n}{S\langle n\rangle}\ln \frac{n}{S\langle n\rangle} + \mathcal{O}(S^{-2})\bigg)\bigg]^S\\
    &\approx \dfrac{1}{S}\Bigg[\frac{S \langle n \rangle- d b  \psi(1+d)- d b  \ln \left(\frac{b}{S \langle n \rangle}\right)}{S \langle n \rangle}\Bigg]^S\\
    &=\dfrac{1}{S}\bigg[1 -\dfrac{\psi(1+d)-\ln(d S)}{S}\bigg]^S\\
    &\overset{S\to \infty}{\longrightarrow} \dfrac{1}{S} e^{-\psi(1+d)+\ln(d S)}\\
    &= e^{-\psi(1+d)}d \ , \label{eq:D-fin}
\end{align}
where $\psi(\cdot)$ is the Digamma function, $\psi(z) = \frac{d}{dz} \ln(\Gamma(z))$.

\begin{figure}
    \centering
    \includegraphics[width=0.7\columnwidth]{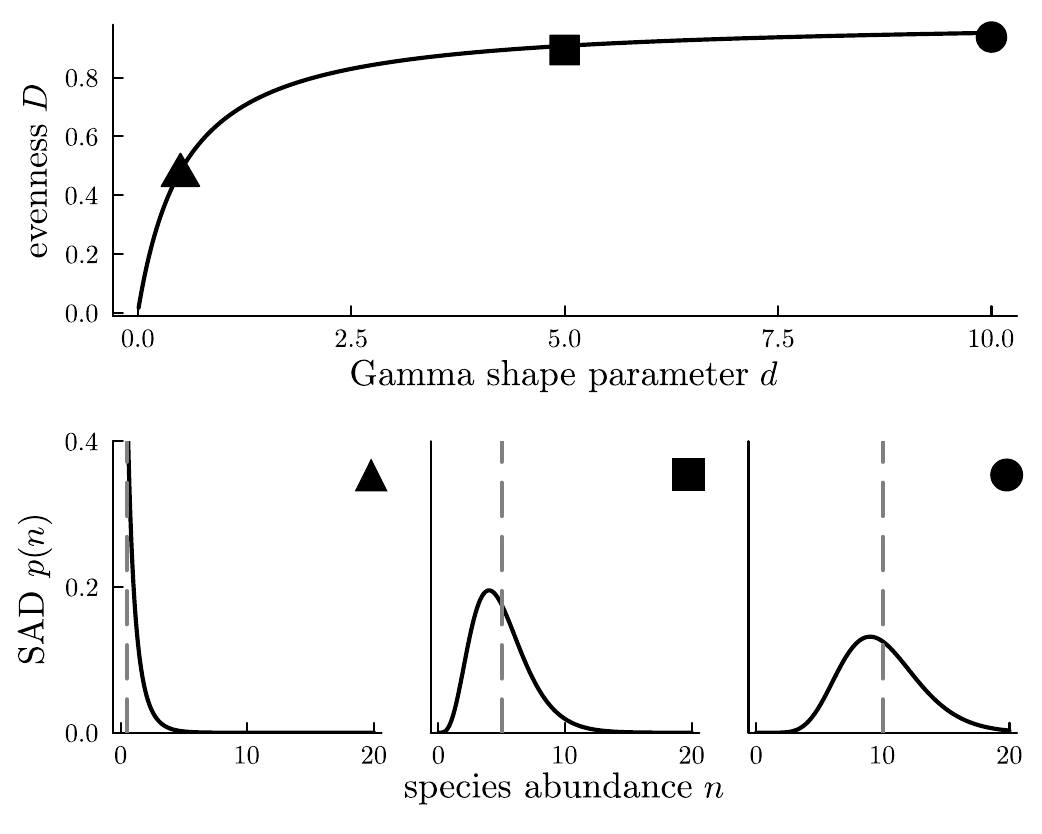}
    \caption{Evenness of different instances of Gamma distributions, $p(n) \propto n^{-1+d}e^{-n/b}$; in all cases, $b = 1$. Dashed lines indicate the mean of the distribution. For large values of $d$, the mean shifts away from $0$, and the coefficient of variation $\sigma(n)/\langle n \rangle = d ^{-\frac{1}{2}}$ tends to 0, so that most abundances sampled from the distribution will have similar values and evenness will be close to one.}
    \label{fig:evenness_gamma_examples}
\end{figure}

\begin{figure}[h!]
    \centering
    \begin{minipage}[c]{0.47\textwidth}
        \centering
        \includegraphics[width=\textwidth]{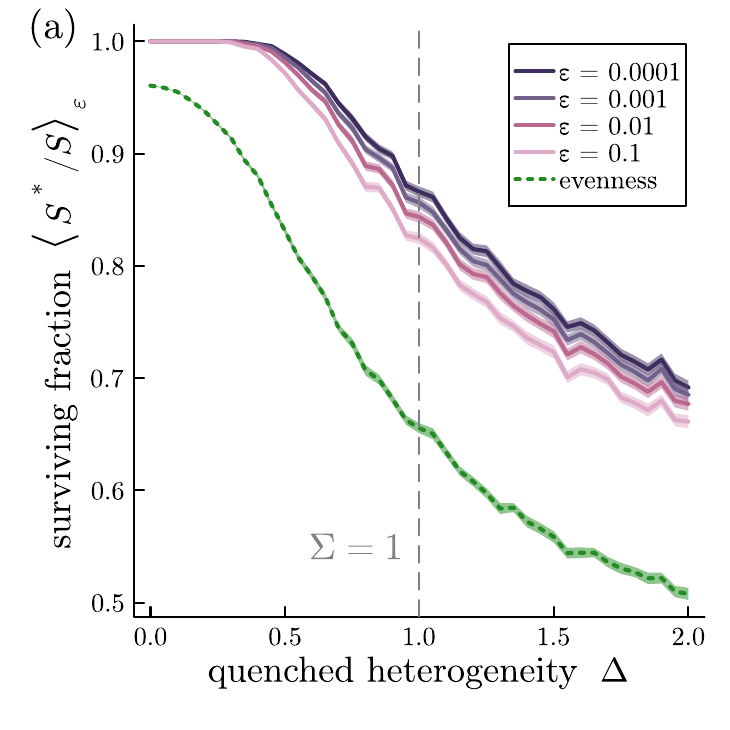}
    \end{minipage}
    \hfill
    \begin{minipage}[c]{0.47\textwidth}
        \centering
        \includegraphics[width=\textwidth]{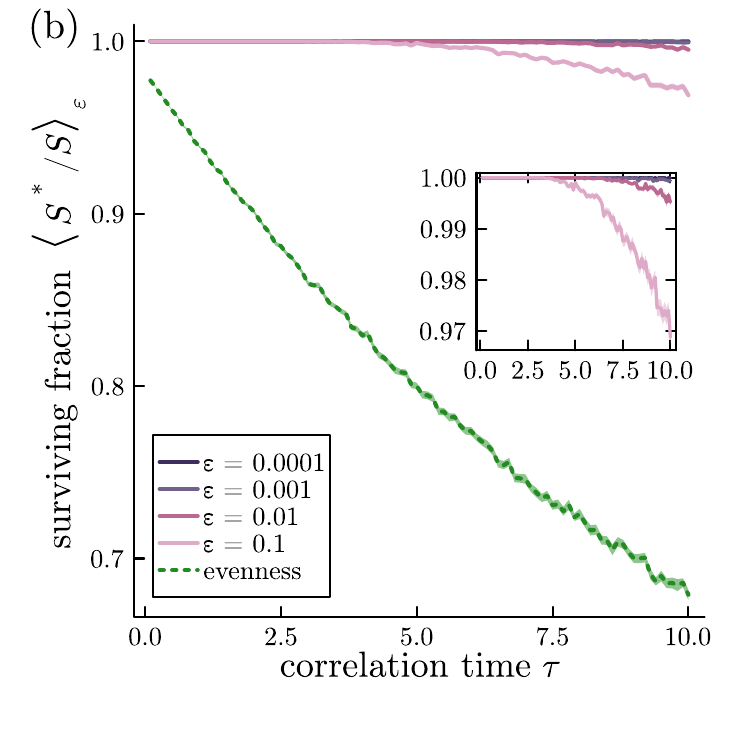}
    \end{minipage}
    \caption{\resub{Fraction of species coexisting at stationarity for different extinction thresholds $\varepsilon$ compared to evenness: lines and envelopes represent mean $\pm$ standard deviation obtained from 100 replicated simulations for each value of $\Delta$ and $\tau$ in panels (a) and (b) respectively. Angular brackets $\left\langle \cdot \right\rangle_{\varepsilon}$ denote averaging over many different realizations where all parameters are fixed, and for a given value of $\varepsilon$. In both figures, $\Sigma = 1$. (a) $\tau = 1$, varying $\Delta$. Notice that, for $\Delta \gtrapprox \Sigma$, $\lim_{\varepsilon \to 0}\left\langle S^*/S \right\rangle_{\epsilon} \neq 1$, where the limit is understood in a numerical sense\finalresub{, as conveyed by the different curves obtained for different values of $\varepsilon$}. Conversely, in the regime $\Delta \ll \Sigma$ (dashed line), $\left\langle S^*/S \right\rangle_{\epsilon} \approx 1$ if $\varepsilon$ is sufficiently close to zero. (b) $\Delta = 0$, varying $\tau$. In this case, $\lim_{\varepsilon \to 0}\left(S^*/S \right)_{\epsilon} = 1$, signaling that there are no true extinction (see the inset for a magnified view of the region of interest). Aside from $\Delta$ and $\tau$, all parameters used in simulations to obtain (a) and (b) are identical (in particular, $\nu = 1$ and $\nu_c = 20$) and are reported in Appendix~\ref{si:numericals}.}}
    \label{fig:CEP_strict_vs_effective}
\end{figure}

Firstly we notice that $D$ is manifestly a scale-free quantity, in that it only depends on the power-law component of the underlying Gamma distribution. Secondly, by an analogous computation to the one shown above, one easily finds that, in the limit $S\to+\infty$, $\langle \big( e^H / S \big)^2\rangle \to D^2$, so that $ e^H / S$ is self-averaging (i.e.,   fluctuations vanish), and in particular $\langle e^H\rangle/S \to e^{\langle H\rangle}/S$. Figure \ref{fig:evenness_gamma_examples} shows how $D$ changes depending on the shape of the underlying Gamma distribution of abundances. \resub{Thirdly, as anticipated in Section~\ref{sec:effective_CEP_violation} of the main text, numerical analysis of the predictions our model, Eqs.~\eqref{eq:model_full}, confirms that evenness bounds the fraction of strictly present species $\left(S^*/S \right)_{\epsilon}$ from below\footnote{\resub{Since any species with vanishing abundance is automatically discarded by $D$, while surviving species contribute according to their abundance by increasing evenness at most by $1/S$, one has $D \leq \left(S^* / S\right)_{\varepsilon}$ for $\varepsilon$ sufficiently close to $0$. \finalresub{Indeed, this is easily shown to hold in general for $\varepsilon = 0$, as we've done in Section~\ref{sec:even}; then, it for $\varepsilon = 0$ the inequality is strict, and assuming that $\left(S^* / S\right)_{\varepsilon}$ depends continuously on $\varepsilon$, it follows that the inequality holds even for $\varepsilon \in [0,\bar\varepsilon]$, for some small enough positive constant $\bar\varepsilon$.} In other words, evenness bounds from below the fraction of species with non-zero abundance; equivalently stated, it always holds $S_{\text{eff}} \leq S^*$, where $S_{\text{eff}} = D \times S$. Thus, the effective breaking of the CEP implies its strict breaking.}}; see Fig.~\ref{fig:CEP_strict_vs_effective}. When extinctions do happen ($\Delta > 0$ in Fig.~\ref{fig:CEP_strict_vs_effective}a), evenness qualitatively matches the information given by $\left(S^*/S \right)_{\epsilon}$; instead, in cases where extinctions are not observed ($\Delta = 0$ in Fig.~\ref{fig:CEP_strict_vs_effective}b), evenness displays a regular pattern which does not depend on any threshold. As a last observation, in the large $S$ limit, we note that we may rewrite $D$ in the equivalent form 
\begin{equation}\label{evenness large S}
    D= e^{\left(\langle n \rangle\ln\langle n\rangle  - \langle n\ln n \rangle\right)/\langle n \rangle}
\end{equation}
 and, thanks to Jensen's inequality applied the the convex function $n\ln n$, we have $D\leq 1$.}

\resub{\subsection{\label{si:evenness_rarefaction} Connection between evenness and species' discovery rate}}

\resub{In the following, we calculate (in the limit of large \finalresub{number of species} $S$) the mean fraction $f$ of species detected by sampling $N$ individuals from the community according to the relative abundances of species. In fact, this quantity is closely related to the \emph{species accumulation curve} (SAC), also known as the ``collector’s curve'' or ``rarefaction curve''~\cite{siegel2004rarefaction}, which describes the cumulative number of species observed as a function of sampling effort. SACs are widely used in ecology to evaluate sampling completeness and biodiversity richness \cite{Ugland2003,Willott2001}. SACs are related to the more commonly known \emph{species–area relationships} (SARs). Although both SACs and SARs relate sampling effort to species richness, they are not equivalent~\cite{Matthews2016} as SARs take into account spatial structure. Additionally, there is a direct connection between SACs and nonparametric richness estimators, particularly those introduced by Ref.~\cite{chao2015estimating}, which use the number of species represented by one or two individuals to infer total species richness of an ecosystem. Our formulation achieves analytical control and complements existing empirical approaches to SACs, while connecting them to the notion of evenness we use throughout this work.

As in the previous section, we consider an ecological community of $S$ species whose SAD is $P(n)$. From the SAD, we obtain the relative abundances, $p_\sigma$, $\sigma = 1,\dots,S$ as in Eq.~\eqref{eq:relative_abundances_SAD}. Let us denote by $Y_\sigma$ the event that species $\sigma$ is sampled at least once in $N$ draws (thus $Y_\sigma \in \{0,1\}$): we find $$\mathbb{E}\left[Y_\sigma \right] =\text{Prob}\left(Y_{\sigma} = 1\right) = 1 - \text{Prob}\left(Y_{\sigma} = 0\right) = 1 - \left(1-p_{\sigma}\right)^N \ .$$ Thus, the (random) number of species detected after $N$ draws is $X_N = \sum_{\sigma=1}^S Y_\sigma$, with expected value $$ \mathbb{E}\left[X_N\right] = \sum_{\sigma=1}^S \mathbb{E}\left[Y_\sigma \right] = \sum_{\sigma=1}^S\left[1 - \left(1-p_{\sigma}\right)^N\right] \ .$$
Now, we compute the fraction of detected species as $f = \left\langle\mathbb{E}\left[X_N\right] / S \right\rangle$, where the angular brackets represent averaging over all possible samples of size $S$ obtained from the SAD. In the limit of large $S$, where each $p_\sigma$ is small, we can expand $(1-p_{\sigma})^N \approx e^{-N p_\sigma} \equiv e^{-\frac{n_\sigma}{\langle n \rangle}N/S}$. Therefore,

\begin{equation}\label{eq:rarefaction_derivation}
f(N/S) = \frac{1}{S}\sum_{\sigma=1}^S\int dn_{\sigma} \ P(n_{\sigma}) \ \left(1 -e^{-\frac{n_\sigma}{\langle n \rangle}N/S}\right) = 1 - \left\langle e^{-\frac{n}{\langle n \rangle}(N/S)} \right\rangle \ ,
\end{equation}
where a trivial multiplication of independent probability measures is omitted, and the sum of identical integrals cancels out with the prefactor $1/S$. Then, by defining the \textit{sampling effort}\footnote{\resub{As argued in Ref.~\cite{Willott2001}, the number $N$ of individuals collected is an appropriate measure of sampling effort. Here, we adopt this very definition of sampling effort up to a rescaling, thus using the continuous value $\lambda = N/S$, in the limit of large number of species $S$.}} $\lambda = N/S$, we find that the SAC is

\begin{equation}\label{eq:detected_fraction_gamma}
    f\left(\lambda\right) = 1 - \Phi\left(-\lambda/\langle n \rangle \right) \ ,
\end{equation}
where $\Phi\left(t\right) = \int dn~P(n)~e^{n t}$ is the moment generating function of $P(n)$. 

$f'(\lambda)$ represents the discovery rate of species during sampling: it holds $0 < f'(\lambda) < 1$, and particularly $f'(0) = 1$ and $\lim_{\lambda \to \infty}f'(\lambda) = 0$. The (unrealistic) perfectly even community, for which the SAD is a Dirac delta function, $P(n) = \delta\left(n - \langle n \rangle\right)$, has exponential discovery rate, namely $f(\lambda) = 1 - e^{-\lambda}$, whereas\footnote{\resub{For ecological relevance, we mention that we can obtain closed or asymptotic forms for the evenness and SACs of other probability distributions which are commonly used as models for SADs, namely power law (Pareto) and LogNormal distributions. Hence, this framework can be employed to analyze different models.}} e.g. for a community with a Gamma SAD with shape parameter $d$ it holds $f(\lambda) = 1 - \left(1+\lambda/d\right)^{-d}$.  More generally, we expect that less even communities will require a greater sampling effort to detect a given fraction of all species comprised in the ecosystem. Then, we can quantify the ``samplability'' of a community by a cost function such as

\begin{equation}\label{eq:samplability}
    \int_0^{+\infty} \left(e^{-\lambda}-f'(\lambda)\right)\mathcal{W}(\lambda)d\lambda ,
\end{equation}
where the positive function $\mathcal{W}(\lambda)$ serves to attribute different weights to discovery rate deviations from the maximally even case at different stages of sampling. The integrand in Eq.~\eqref{eq:samplability} is either identically equal to 0 (this happens when $f'(\lambda) = e^{-\lambda}$, i.e. for a Dirac delta SAD), or becomes negative for $\lambda > \bar\lambda$, where $\bar\lambda$ depends on the SAD. Hence, we may interpret the integral in Eq.~\eqref{eq:samplability} as the difference of the two contributions $\int_0^{+\infty} e^{-\lambda}\mathcal{W}(\lambda)d\lambda$ and $\int_0^{+\infty} f'(\lambda)\mathcal{W}(\lambda)d\lambda$, for $\mathcal{W}(\lambda)\propto \lambda^{-1+p}$ and $p > 0$. With this choice, both contributions are finite and positive; moreover, the difference of these two integrals is finite even in the limit $p \to 0^+.$

Crucially, evenness is found to be a decreasing function of exactly the quantity defined in Eq.~\eqref{eq:samplability}, with $\mathcal{W}(\lambda) = \lambda^{-1}$:

\begin{equation}\label{eq:evenness_vs_samplability}
D = \exp\left(-\int_0^{+\infty} \frac{e^{-\lambda}-f'(\lambda)}{\lambda}d\lambda\right) \ .
\end{equation}
To prove Eq.~\eqref{eq:evenness_vs_samplability}, we begin by rewriting evenness in Eq.~\eqref{evenness large S} as

\begin{equation}\label{eq:evenness_nlogn}
    D = \langle n \rangle e^{-\frac{\left\langle n \ln n \right\rangle}{\langle n \rangle}} \ .
\end{equation}
We make use of a Frullani identity,

\begin{equation}\label{eq:frullani_integral}
    \ln n = \int_{0}^{+\infty}\frac{e^{-t}-e^{-nt}}{t}dt \ ,
\end{equation}
so as to rewrite

\begin{equation}\label{eq:frullani_nlogn_avg}
    \left\langle n \ln n\right\rangle = \int_{0}^{+\infty}\frac{\langle n \rangle e^{-t}-\langle n  e^{-n t}\rangle}{t}dt \ .
\end{equation}
In the integral in Eq.~\eqref{eq:frullani_nlogn_avg}, we recognize the term $$\left\langle n  e^{-n t}\right\rangle = -\frac{d}{dt}\Phi(-t) = \frac{d}{dt}\left(1 - \Phi(-t)\right) \ .$$ From this, first with the change of variables $\lambda = t \langle n \rangle$ and then with simple algebraic manipulations, we obtain 

\begin{equation}\label{eq:frullani_steps}
    \frac{\left\langle n \ln n \right\rangle}{\langle n \rangle} = \int_{0}^{+\infty}\frac{e^{-\lambda / \langle n \rangle}-f'(\lambda)}{\lambda}d\lambda = \ln \langle n \rangle + \int_{0}^{+\infty}\frac{e^{-\lambda}-f'(\lambda)}{\lambda}d\lambda \ .
\end{equation}
Plugging the last member of Eq.~\eqref{eq:frullani_steps} in Eq.~\eqref{eq:evenness_nlogn}, we immediately obtain Eq.~\eqref{eq:samplability}.

Clearly, if $f(\lambda) = 1 - e^{-\lambda}$ (corresponding to a Dirac delta SAD), evenness in Eq.~\eqref{eq:evenness_vs_samplability} assumes its maximum value, $D = 1$, while in all other cases $D < 1$. Thus, given the explicit connection between evenness and species' discovery rate given by Eq.~\eqref{eq:samplability}, we can interpret evenness as quantifying how quickly species are discovered when uniformly sampling individuals from a well-mixed ecosystem, with highly even communities requiring the least effort to gauge the underlying species richness.}

\renewcommand{\thefigure}{E\arabic{figure}}
\setcounter{figure}{0}

\section{Additional analysis for $\Delta = 0$}\label{si:local_time}

\begin{figure}[h!]
    \centering
    \includegraphics[width=0.9\textwidth]{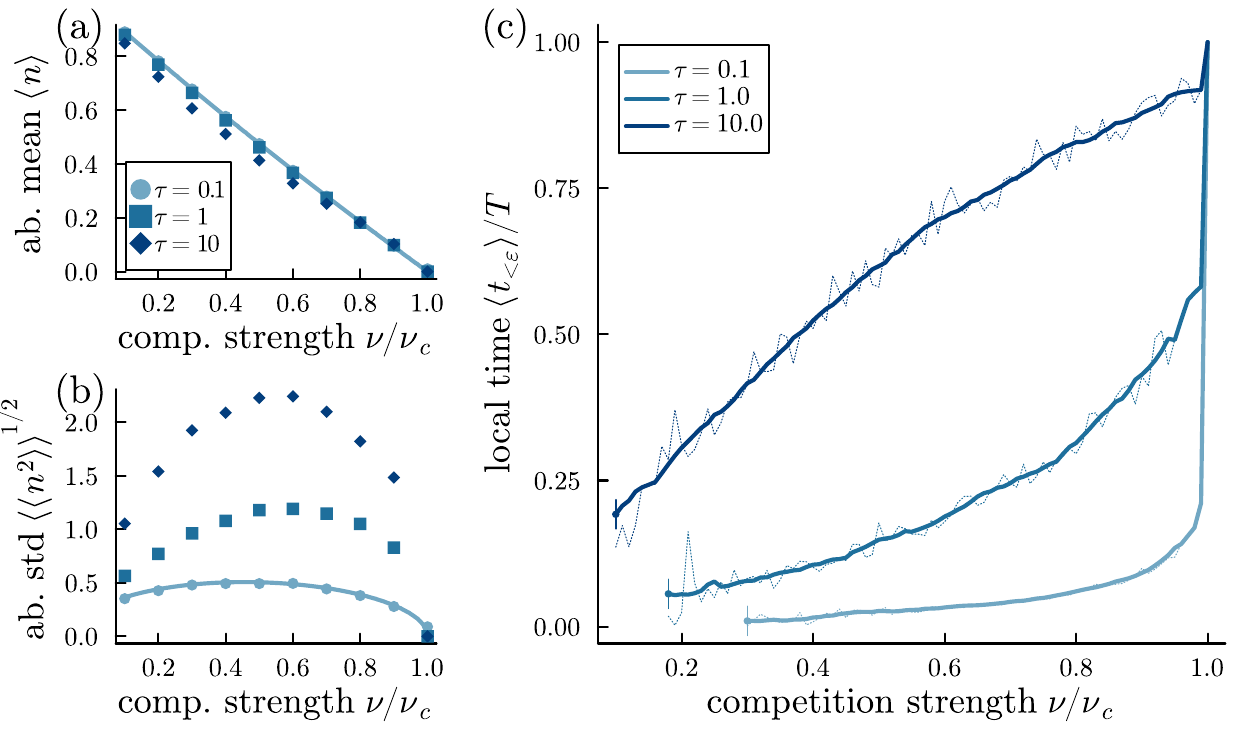}
    \caption{Comparing the effect of different correlation times on stationary statistics. Mean (a) and standard deviation (b) of stationary species abundances for different values of $\nu$. Markers are empirical values obtained from simulations, while solid line is the white noise limit analytic prediction. (c) Mean local time $\langle t_{< \varepsilon}\rangle$, i.e., mean length of time intervals spent by any species below a given threshold $\epsilon$ (in this case, $\epsilon = 10^{-3}$), computed numerically; thick lines are smoothed (i.e. locally averaged) results, while thin dotted lines display typical fluctuations. As $\nu \to \nu_c$, mean local time saturates the observation time $T$, i.e., it diverges in the limit $T \to +\infty$, corresponding to the fact that all species become extinct at stationarity. Notice that different lines, representing mean local times for different values of $\tau$, begin at different values of $\nu/\nu_c$. This signals that, for $\nu \ll \nu_c$ and low $\tau$, no species will pass through the selected threshold. Correspondingly, we expect the peak of the SAD to move away from $n = 0$.}
    \label{fig:coeffvar_trajectories}

\end{figure}

For $\Delta = 0$, we obtain complementary information to that of the SAD by the direct examination of statistics of species abundances, shown in Fig.~\ref{fig:coeffvar_trajectories}, for increasing competition strength $\nu / \nu_c$ and for different values of $\tau$. In all cases, as $\nu / \nu_c \to 1$, the mean abundance of species decreases to $0$ linearly, while variance changes in a non-monotonic fashion. Increasing the correlation time affects mean abundances only marginally, as its linear decrease with competition strength is qualitatively maintained. Conversely, the maximum in variance becomes more pronounced with increasing $\tau$, showing a significant departure from the white noise limit result and formation of long tails in the distribution of abundances. 

Concurrently, an analysis of temporal statistics of species abundances shows that the length of time intervals that species spend below a threshold, which can be interpreted as \textit{detectability} threshold, diverges as $\nu$ approaches $\nu_c$ more quickly for larger values of $\tau$. This means that, while in all cases increasing competition strength will result in lower evenness, in the presence of rapid fluctuations most species will easily grow from rare to relatively abundant, even in competitive ecosystems; on the other hand, if environmental fluctuations play out on longer time scales, rare species will tend to remain so.

\begin{figure}[h!]
\begin{center}
        \centering
        \includegraphics[width=0.7\textwidth]{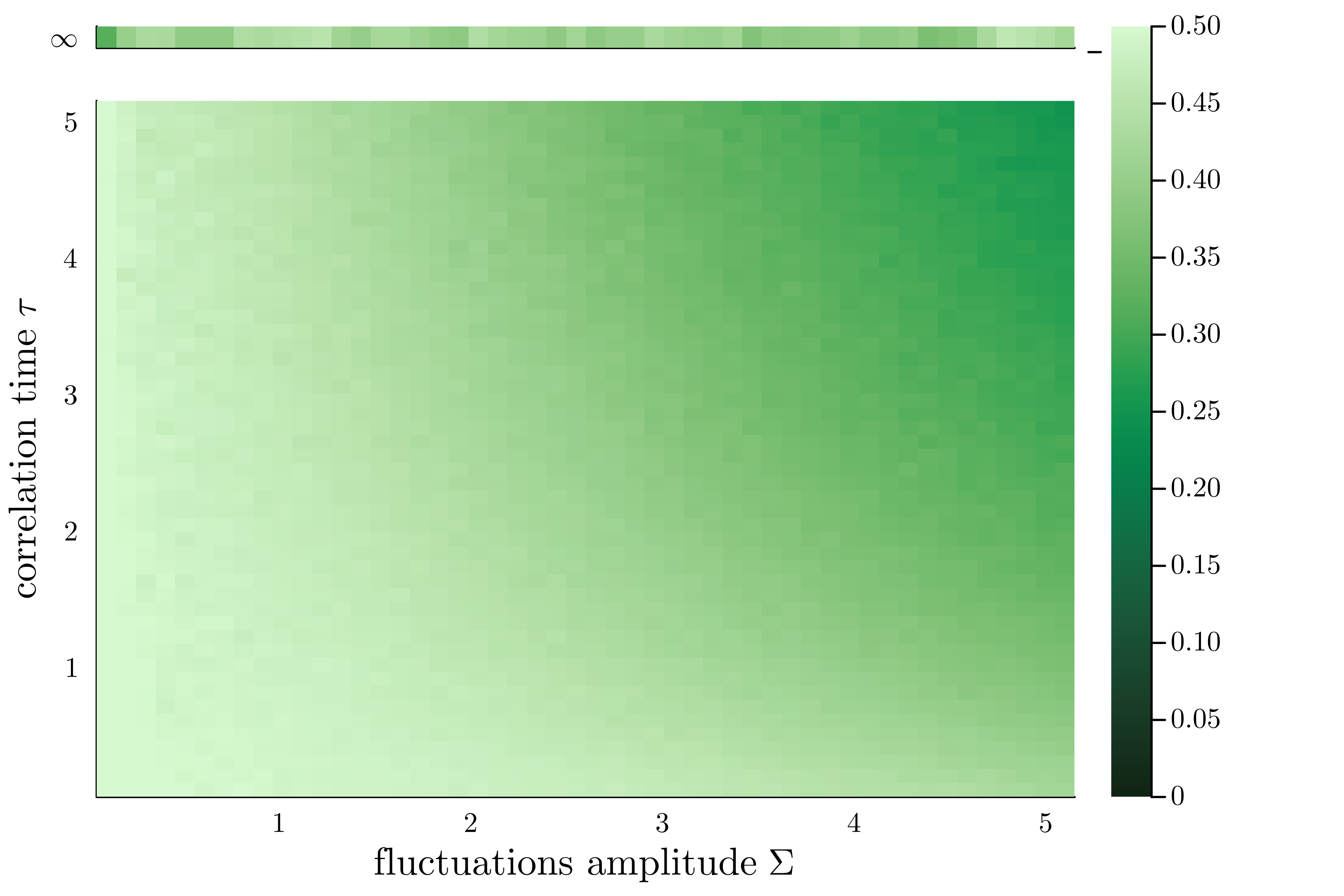}
\end{center}
\caption{Evenness as function of $\Sigma$ and $\tau$, for $\Delta = 0$; $\nu = 1$. Each pixel is the result of a single realization of the system; the indicated value of evenness has been obtained by time average of the time dependent evenness, after discarding transients. The top line represents the corresponding values of evenness for the specified $\Sigma$ in the quenched limit, here realized with $\tau = 10^{100} \approx \infty$. Notice that in the quenched limit evenness fluctuates widely, reflecting how the heterogeneity of steady state community structure depends on the realization of the system for relatively low $S$ and $R$. All parameters are reported in Appendix~\ref{si:numericals}.}
\label{fig:phase_diagrams_annealed}
\end{figure}

In terms of evenness, we observe no sharp transition between states of higher and lower evenness; see Fig.~\ref{fig:phase_diagrams_annealed}. We obtain a continuous gradient, as opposed to the case of $\Delta > 0$.

% also here, using a PNG image to circumvent an issue between Julia's Plot.jl and some Apple devices

\section{Stationary distributions for the case of abiotic resources}\label{si:abiotic}

In general one might be interested in considering different types of dynamics for resources, in order to better represent different ecosystems. While stationary distribution of resources will change with its dynamics, the species abundance distribution will have the same form. As an important example, we consider abiotic dynamics for resources, namely the modification to Eqs.~\eqref{eq:model_full}

\begin{equation}
\dot{c}_i = \mu c_i (1 - c_i / k) + \dots \rightarrow \dot{c}_i = \phi - \gamma c_i + \dots
\end{equation}
The derivation of effective equations is identical to the case already considered, as is the study of stationary distributions for consumers and resources, which are identical in form to those in Eq.~\eqref{eq:wnl_SAD}. However the coefficients of $p_{st}(c)$ are redefined to be $d_c = -2\frac{\gamma+\bar\alpha\langle n \rangle}{\Sigma^2 \langle n^2 \rangle}$, $b_c = \frac{\langle n^2 \rangle}{\langle n \rangle}$ and $l_c = 2 \frac{\phi + \lambda_c}{\Sigma^2 \langle n^2 \rangle}$. Notice that $d_c < 0$, while $l_c > 0$ even in the case $\lambda_c = 0$: this distribution is normalizable with $\mathcal{Z}_c^{-1} = 2(b_c l_c)^{d/2}K_{d}\Big(2\sqrt{l/b}\Big)$, where $K_\cdot(\cdot)$ is a modified Bessel function of the second kind. In fact the first two moments of the distributions can be written as $\langle c \rangle = \sqrt{b_c l_c}\frac{K_{1+d_c}\big(2\sqrt{l_c/b_c} \big)}{K_{d_c}\big(2\sqrt{l_c/b_c}\big)}$ and $\langle c^2 \rangle = b_c l_c\frac{K_{2+d_c}\big(2\sqrt{l_c/b_c} \big)}{K_{d_c}\big(2\sqrt{l_c/b_c}\big)}$. In this case, self-consistent equations may only be solved numerically.

\section{Details on simulations and analysis}\label{si:numericals}

All code used to perform numerical analysis was written in \texttt{Julia 1.9.3}. All figures have been produced through the \texttt{Plot.jl} package, using the default \texttt{GR} backend. To fit curves to points, we used the function \texttt{LsqFit.curve\_fit}, while to build hisogram from accumulated data we used \texttt{Distributions.fit} to populate an \texttt{Histogram} object. The random numbers used to build the trajectories of metabolic strategies have been generated through the function \texttt{Base.randn()}. Numerical integration of Eqs.~\ref{eq:model_full} was performed through custom code, without using \texttt{DifferentialEquations.jl} or similar packages.  

For $\tau > 0$, we used the \textit{Euler-Maruyama} scheme~\cite{gardiner2004handbook} to integrate the equations of motion, taking cautions that the integration step $dt$ is smaller than the correlation time $\tau$ by at least an order of magnitude, and in any case not larger than $10^{-2}$. While choosing smaller values of $\tau$ well approximates the white noise results, a rigorous analysis of the case $\tau = 0$ requires that numerical integrations uses \textit{exactly} white noise; in this limit, it is well known that one must use Stratonovich calculus~\cite{vankampen1992}, which is implemented numerically by Heun's scheme~\cite{ruemelin1982}. In this case, choosing $dt = 10^{-1}$ is often sufficient to obtain accurate results; however, as $\nu \to \nu_c$, smaller time steps are required to avoid abundances reaching 0 in finite times. Similarly, probing large values of $\Sigma$ typically required smaller time steps, $dt \lessapprox 10^{-3}$.

Initial conditions of both $n_{\sigma}$ and $c_{i}$ have been sampled by uniform distributions with support on positive real numbers, but stationary distributions has been found to be independent of initial conditions. Initial conditions of $Z_{\sigma,i}$ have been sampled from the stationary distribution of the corresponding Ornstein-Uhlenbeck process. Without this precaution, and especially in simulations for which $\tau \gtrapprox 10 $ or $\nu \to \nu_c$, the time required for the system to reach the steady state may become exceedingly large.

To obtain numerically the stationary distributions to which compare those of the effective process, Eqs.~\eqref{eq:DMFT_effective_dynamics}, we simulated Eqs.~\eqref{eq:model_full} in a time interval $[0,t_{\text{max}}]$, and accumulated all values of $n_\sigma$ and $c_i$, after discarding all abundances in the transient time interval $[0,f t_{\text{max}}]$, with $0 < f < 1$. The choice of both $t_{\text{max}}$ and $f$ has been made on the empirical basis: as a thumb rule, larger values of $\tau$ lead to longer transients times to go from the initial distribution to the steady state distributions. Typically, for white noise simulations, we took $t_{\text{max}} = 10^{3}$ and $f = 0.5$, while for $\tau \gtrapprox 10$ we set $t_{\text{max}} = 10^{4}$ and $f = 0.7$. Furthermore, for large values of $\tau$, and due to long-lasting fluctuations, larger values of $S$ and $R$ need to be chosen, in order to adequately populate the empirical steady state distribution, so that comparisons with the analytical results are meaningful.

\subsection*{Figures: parameters used and additional notes}

\noindent \textbf{Fig.~\ref{fig:trajectories_examples}} Fixed parameters are $\bar\alpha = 5$, $\Sigma = 1$, $\Delta = 0$, $\delta = 1$, $\mu = 10$, $\kappa = 1$, $S = 125$, $R = 50$, $dt = 10^{-2}$, $t_{\text{max}} = 200$.\\

\noindent \textbf{Fig.~\ref{fig:SAD_wnl_ucna}} (a) Fixed parameters are $\bar\alpha = 10$, $\Sigma = 1$, $\Delta = 0$, $\delta = 1$, $\mu = 10$, $\kappa = 1$, $R = 50$, $dt = 10^{-1}$, $t_{\text{max}} = 5000$, $f = 0.5$. Simulations have been carried out directly with white noise, i.e. not by choosing a value of $\tau$ close to 0.
(b) Fixed parameters are $\bar\alpha = 10$, $\Sigma = 1$, $\Delta = 0$, $\delta = 1$, $\mu = 10$, $\kappa = 1$, $R = 500$, $dt = 10^{-1}$, $t_{\text{max}} = 10^{4}$, $f \in [0.5,0.8]$ depending on $\tau$. Additionally, we report that, for large $\tau$, in order to successfully fit the theoretical distribution to empirical data, a suitable cut-off $\epsilon$ have to be applied to abundances, meaning that abundances lower than $\epsilon$ have not been accumulated to form empirical steady state distributions. For example, for $\tau = 100$, we used $\epsilon = 0.05$. The fit of the theoretical distribution has been performed using as free parameters both the constants $\chi_{x}$, $x=n,c$, as explained in the main text, and the normalization constants of the distributions.\\

\resub{\noindent \textbf{Fig.~\ref{fig:D_vs_Sigma_tau}} (a) Fixed parameters are $\bar\alpha = 20$, $\Delta = 0$, $\delta = \kappa = 1$, $\mu = 100$, $t_{\text{max}} = 500$, $dt = 10^{-3}$, $f = 0.5$ $R = 10$.\\ (b) Fixed parameters are $\bar\alpha = 20$, $\Delta = 0$, $ \delta = \kappa = 1$, $\mu = 100$, $t_{\text{max}} = 5000$, $dt = 10^{-3}$, $f = 0.5$ $S = R = 100$.}\\

\noindent \textbf{Fig.~\ref{fig:evenness_vs_nu}} (a) Fixed parameters are $\bar\alpha = 10$, $\Delta = 0$, $\Sigma = \delta = \kappa = 1$, $t_{\text{max}} = 1000$, $dt = 10^{-3}$, $f = 0.7$ $R = 100$.\\ (b) Fixed parameters are $\bar\alpha = 10$, $\Delta = 0$, $\Sigma = \delta = \kappa = 1$, $\mu = 100$, $t_{\text{max}} = 1000$, $dt = \min\big\{10^{-2},\tau/10\big\}$, $f = 0.7$ $R = 500$.\\

\noindent \textbf{Fig.~\ref{fig:sad_quenched}} Fixed parameters are $\bar\alpha = 10$, $\Sigma = \delta = \kappa = 1$, $\mu = 100$, $\tau = 0.1$, $t_{\text{max}} = 10^3$, $dt = 10^{-2}$, $f = 0.5$, $S = R = 500$. (b) Fixed parameters are $\bar\alpha = 50$, $\Delta = 0.1$, $\delta = \kappa = 1$, $\mu = 100$, $t_{\text{max}} = 5000$, $dt = 10^{-2}$, $f = 0.5$, $S = R = 50$.\\ 

\resub{\noindent \textbf{Fig.~\ref{fig:CEP_strict_vs_effective}} (a) Fixed parameters are $\bar\alpha = 20$, $\tau=0.5$, $\Sigma = \delta = \kappa = 1$, $\mu = 10$, $t_{\text{max}} = 1000$, $dt = 10^{-2}$, $f = 0.9$, $S = R = 50$. (b) Fixed parameters are $\bar\alpha = 20$, $\Delta=0$, $\Sigma = \delta = \kappa = 1$, $\mu = 10$, $t_{\text{max}} = 1000$, $dt = 10^{-2}$, $f = 0.9$, $S = R = 50$. For both (a) and (b), results' statistics (lines are means, envelopes are standard deviations) are obtained through 100 replicates of simulations.} \\

\noindent \textbf{Fig.~\ref{fig:coeffvar_trajectories}} Fixed parameters are $\bar\alpha = 10$, $\delta = \kappa = 1$, $\mu = 10$. $t_{\text{max}} = 1000$, $dt = 10^{-2}$, $f = 0.5$, $R = 50$.\\

\noindent \textbf{Fig.~\ref{fig:phase_diagrams_annealed}} Fixed parameters are $\bar\alpha = 50$, $\Delta = 0$, $\delta = \kappa = 1$, $\mu = 10$, $t_{\text{max}} = 5000$, $dt = 10^{-2}$, $f = 0.5$, $S = R = 50$.

\twocolumngrid

\bibliography{mybib}{}
\bibliographystyle{unsrt}

\end{document}